\documentclass[twocolumn,onecolappendix]{aastex61}
\usepackage{natbib,aas_macros}
\usepackage{graphicx}	
\usepackage{amsmath}	
\usepackage{amssymb}	
\usepackage{bm}		    
\usepackage{xspace} 
\usepackage{breakurl}

\newcommand{\BVI }{\rm{$B,V,I_c$}}
\newcommand{\JHK }{\rm{$J,H,K_s$}}
\newcommand{\LM }{\rm{[3.6] \& [4.5]}}

\newcommand{\gloess}{GLOESS}
\newcommand{\spitzer}{\emph{Spitzer}}
\newcommand{\hst}{\emph{HST}}
\newcommand{\gaia}{\emph{Gaia}}
\newcommand{\hip}{\emph{Hipparcos}}
\newcommand{\wise}{\emph{WISE}}

\begin{document}

\title[Classical Galactic Field RR Lyrae]{Standard Galactic Field RR Lyrae. I. Optical to Mid-Infrared Phased Photometry}

\author{Andrew~J.~Monson} 
\altaffiliation{The Observatories of the Carnegie Institution for Science, 813 Santa Barbara St., Pasadena, CA 91101, USA}
\altaffiliation{Department of Astronomy \& Astrophysics, The Pennsylvania State University, 525 Davey Lab, University Park, PA 16802, USA}

\author{Rachael~L.~Beaton}
\altaffiliation{The Observatories of the Carnegie Institution for Science, 813 Santa Barbara St., Pasadena, CA 91101, USA}

\author{Victoria~Scowcroft}
\altaffiliation{The Observatories of the Carnegie Institution for Science, 813 Santa Barbara St., Pasadena, CA 91101, USA}
\altaffiliation{Department of Physics, University of Bath, Claverton Down, Bath BA2 7AY, UK}
\altaffiliation{50th Anniversary Prize Fellow}

\author{Wendy~L.~Freedman}
\altaffiliation{Department of Astronomy \& Astrophysics, University of Chicago, 5640 South Ellis Avenue, Chicago, IL 60637, USA}

\author{Barry~F.~Madore}
\altaffiliation{The Observatories of the Carnegie Institution for Science, 813 Santa Barbara St., Pasadena, CA 91101, USA}

\author{Jeffrey~A.~Rich}
\altaffiliation{The Observatories of the Carnegie Institution for Science, 813 Santa Barbara St., Pasadena, CA 91101, USA}

\author{Mark~Seibert}
\altaffiliation{The Observatories of the Carnegie Institution for Science, 813 Santa Barbara St., Pasadena, CA 91101, USA}

\author{Juna~A.~Kollmeier}
\altaffiliation{The Observatories of the Carnegie Institution for Science, 813 Santa Barbara St., Pasadena, CA 91101, USA}

\author{Gisella Clementini}
\altaffiliation{INAF-Osservatorio Astronomico di Bologna, via Ranzani 1, I-40127, Bologna, Italy}

 

\begin{abstract}
We present a multi-wavelength compilation of new and previously published  photometry for 55 Galactic field RR Lyrae variables. 
Individual studies, spanning a time baseline of up to 30 years, 
 are self-consistently phased to produce light curves in 10 photometric bands covering the
 wavelength range from 0.4 to 4.5 microns. 
Data smoothing via the \gloess{} technique is described and applied to generate high-fidelity light curves, from which
  mean magnitudes, amplitudes, rise-times, and times of minimum and maximum light are derived.
60,000 observations were acquired using  the new robotic Three-hundred MilliMeter Telescope (TMMT), which was first deployed at the Carnegie Observatories in Pasadena, CA, and is now permanently installed and operating at Las Campanas Observatory in Chile.
We provide a full description of the TMMT hardware, software, and data reduction pipeline. 
Archival photometry contributed approximately 31,000 observations. 
Photometric data are given in the standard 
  Johnson $UBV$, Kron-Cousins $R_CI_C$, 2MASS JHK, and {\it Spitzer} [3.6] and [4.5] bandpasses.
\end{abstract}


\smallskip
\section{Introduction} \label{sec:intro}

RR Lyrae variables (RRL) are evolved, low-metallicity, He-burning variable stars.
They are extremely important for distance determinations because at infrared wavelengths
 their period-luminosity relationship shows incredibly small scatter in 
 Galactic and LMC clusters \citep{longmore_1986,longmore_1990,dallora_2004,braga_2015}.
At near-infrared wavelengths the scatter can be as low as $\sigma=0.02$~mag, which translates
to  $\sim$~1\% uncertainty in distance to an individual star \citep[see detailed discussion in][]{beaton_2016}.

While the body of work in Galactic clusters sets the foundation for exquisite differential 
 distances using the RRL period-luminosity-metallicity (PLZ) relation  and yields well-defined slopes, 
 a direct calibration of zero-point, slope, and metallicity parameters using geometric 
 distance estimates has been elusive since their discovery over a century ago \citep[based on the work of Williamina Fleming as published in][]{pickering_1901}\footnote{{ The introduction to \citet{smith_1995} also provides a detailed history of the discovery of RR Lyrae among the variable sources discovered in globular clusters in the late 19th century.}}.
While there were over 100 RRL in the \hip{} catalog, RR Lyr itself was the only variable of this class that was both sufficiently bright and near enough to determine a parallax with an uncertainty
 of less than 20 per cent \citep{perryman_1997,vanleeuwen2007}. 
Later, \citet{benedict_2011} derived trigonometric parallaxes using the Fine Guidance Sensor aboard the \hst (HST) for five field RRL with individual quoted uncertainties at the level of 5\%-10\%.  
While the work of \citet{benedict_2011} did provide the first truly geometric foundation, a relatively small sample size still limits the overall statistical accuracy and is not necessarily an improvement over PLZ determinations using Local Group objects with distances independently derived by other means (e.g., using star clusters and main-sequence fitting or dwarf galaxies with precise distances derived by other techniques, such as eclipsing binaries).  
The \gaia{} mission \citep{isgaia} is poised to provide the first opportunity for such a measurement (geometrically based with a large sample size) and it is our purpose in this and related works to provide the necessary data to make full use of the highest-precision \gaia{} RRL sample.

In this work, we present optical and infrared data for 55 of the nearest 
 and brightest Galactic field RRL stars.
These stars span $7.5 < \langle V \rangle < 13.4$ mag in $V$ magnitude, with the majority of them falling 
in the magnitude range for which { \gaia{} is expected to provide} trigonometric parallaxes with a precision better than $\sim$10 microarcseconds ($\mu$as) 
 \citep[see Table 1 in][for predicted end-of-mission values { that will likely be updated with the first Gaia-only parallaxes in DR2}]{debruijne_2014}.
{ These 55 RRL were selected as part of the Carnegie RR Lyrae Program (CRRP), which has the primary goal of establishing the foundation for a Population II-based distance scale utilizing the near- and mid-infrared properties of RR Lyrae stars in the Local Group.
This work supports both the Carnegie-Chicago Hubble Program \citep[CCHP; an overview is given in][]{beaton_2016}, aiming to produce a completely Population II extragalactic distance scale, and the \emph{Spitzer} Merger History and Shape of the Galactic Halo (SMHASH; V. Scowcroft et al. 2017 in preparation), aiming to construct precision three-dimensional maps of the Population II-dominant portions of our Galaxy (e.g., the bulge and stellar halo).
These stars were selected to span a large range of metallicity, have low Galactic extinction, have a moderate incidence of Blazhko stars, and have maximum overlap with other distance measurement techniques like the Baade-Wesselink method \citep{baade_1926,wesselink_1946}\footnote{A detailed introduction of this method is given in Section 2.6 of \citet{smith_1995}.}.}

The organization of the paper is as follows.
{ A detailed discussion of the properties of the CRRP RRL sample is given} in Section \ref{sec:sample}. 
Our hardware and targeted optical-monitoring campaign are described in Section \ref{sec:tmmt}. 
Archival studies are described in Section \ref{sec:data}.
The procedures adopted for phasing individual data sets are described in Section \ref{sec:datamerge}.
{ Our algorithm (\gloess{}) to determine mean magnitudes and provide uniformly sampled light curves is described and applied} in Section \ref{sec:meanmags}. 
Finally, a summary of this work is provided in Section \ref{sec:sum}.
{ Appendix \ref{app:img_prox} gives the technical details for processing data from our custom hardware
 and Appendix \ref{app:indstars} provides detailed information for each star in our sample.}

\begin{figure*}
  \includegraphics[width=\textwidth]{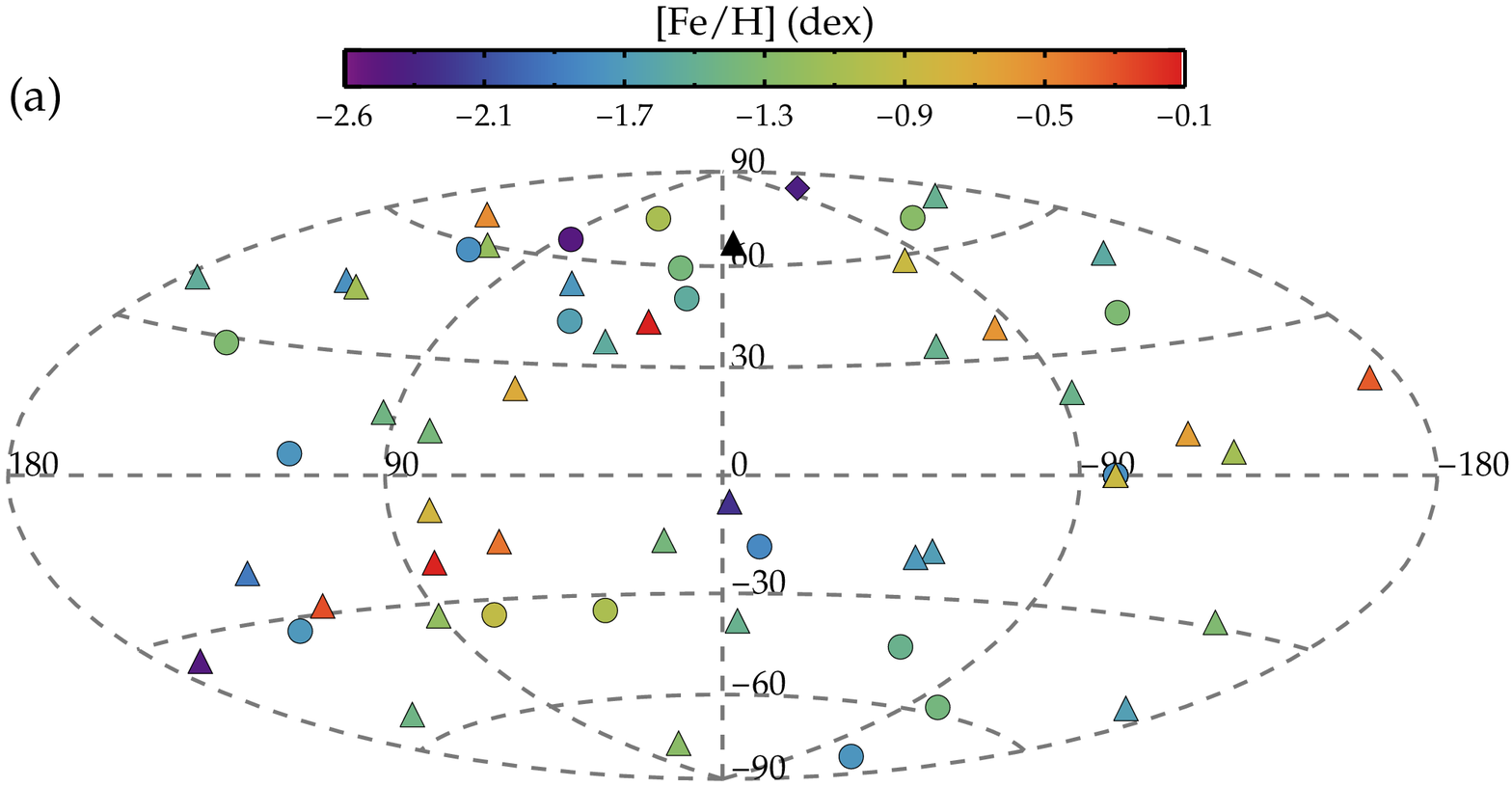}
  \includegraphics[width=\textwidth]{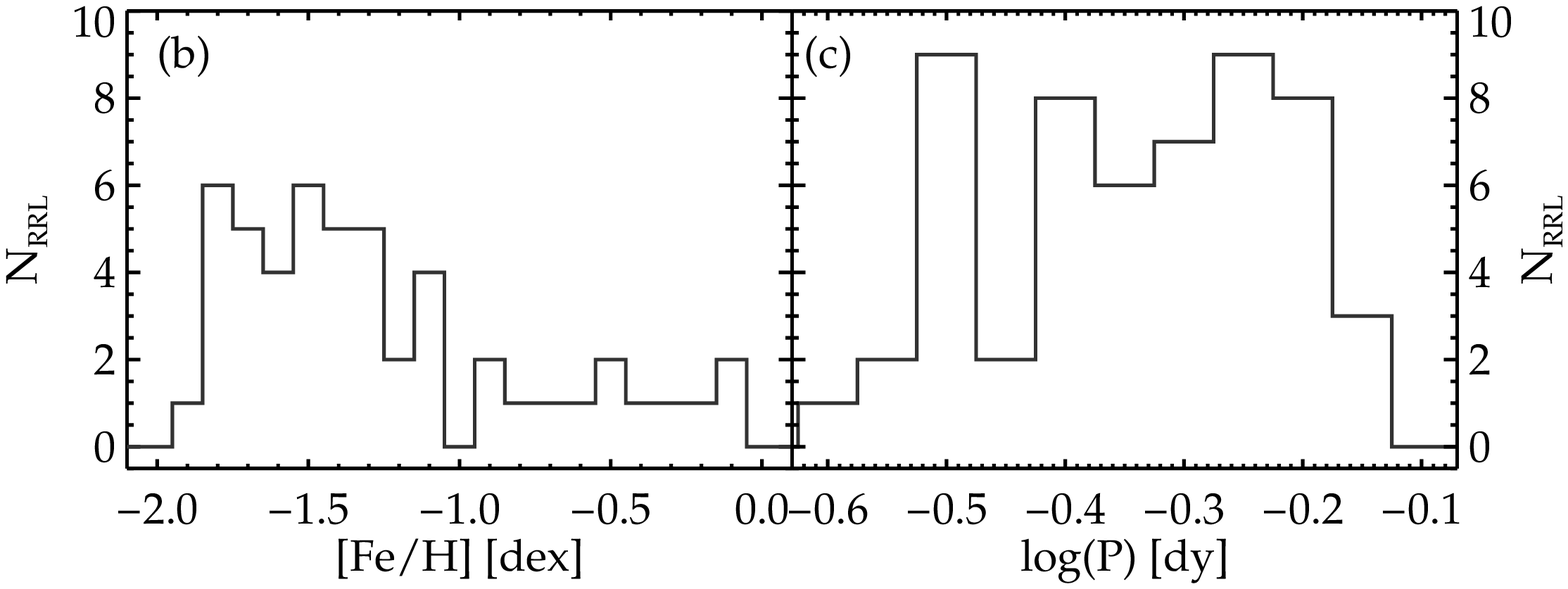}
  \caption{\label{fig:rrldist}
   The sample of RR Lyrae Galactic Calibrators.
   { (a)} Distribution of targets in Galactic latitude ($l$) and longitude ($b$) 
     color-coded by metallicity. 
    Stars of type RRab are shown as triangles, RRc as circles, and RRd as a diamond. 
    The target RR Lyrae are largely out of the Galactic plane ($|b| > 10^{\circ}$), 
     where variations in line-of-sight extinction due to dust variations 
     are minimal \citep[e.g.,][]{zasowski_2009}.
   {\bf (b)} Marginal distribution of $\text{[Fe/H]}$ for our targets, 
     demonstrating that the metal-poor end is well populated,
     but there are stars forming a high-metallicity tail.
   {\bf (c)} Marginal distribution of $\log(P)$ for our targets, 
     emphasizing a relatively uniform distribution.}
\end{figure*}

\section{The RR Lyrae Calibrator Sample} \label{sec:sample}
In this section we describe the demographics of our sample of 55 stars; a summary of the sample properties is given in Table~\ref{tab:sample}.
A comprehensive introduction to RRL is given by \citet{smith_1995}, with an updated presentation provided by \citet{catelan_2015}. 
As we describe our sample we briefly summarize the basics of RRL as needed for the purposes of this paper.

\subsection{Demographics of Pulsation Properties}
There are two primary subtypes of RRL, 
 those that pulsate in the fundamental mode (RRab) and 
 those that pulsate in the first overtone mode (RRc; i.e., there is a pulsational node \emph{within} the star). 
A third subtype pulsates in both modes simultaneously --- these are known as RRd-type variables. 
Generally, for stars of the same density, 
 the ratio of the fundamental period ($P_{F}$) and the first-overtone period ($P_{FO}$)
 is approximately $P_{FO}$/$P_{F}$ = 0.746 (or $\Delta[\log(P)] = 0.127$) 
 and for many purposes the first overtone period can be converted to the 
 equivalent fundamental period using this relationship. 
 Application of this shift is often termed  ``fundamentalizing the period.''
While RRab stars on average have longer periods than their first-overtone counterparts,
 the classification of a given star is generally based on the shape and amplitude of the optical light curve,
  with RRab stars having, { in general, larger amplitudes} and an asymmetric, 
  `saw-tooth' shape and RRc stars looking significantly more symmetric and sinusoidal.
The distribution of RRL into these classifications is correlated with the specific (color/temperature) location
 of the star within the instability strip, 
 which in turn is determined by where the star lands on the horizontal branch based on its post He-flash stellar structure and mass loss. 
{ RRc are hotter (bluer) than RRab, and the RRd-type stars fall in the color/temperature transition region between these two groupings.}
Based on their published classifications in the literature, 
 37 stars in our sample pulsate in the fundamental mode (RRab, 67\%), 
 17 are in the first-overtone mode (RRc, 31\%), 
 and one star, CU Com, is a double-mode pulsator (RRd, 2\%). 

 While many individual RRL have stable pulsation properties, 
 as a population they can show modulation effects in the shapes and amplitudes of their light curves and in their periods.
Light-curve modulations range both in their amplitudes --- some even at the millimagnitude 
 \citep[only detectable with the exquisite time resolution 
 and photometric stability for projects like \emph{Kepler} in space and OGLE from the ground; see][respectively]{nemec_2013, smolec_2015} 
 and some at the level of tenths of a magnitude \citep[e.g., those discussed in][]{smith_1995}---and in their timescales --- some with very short few-day periods \citep[e.g.,][]{nemec_2013} 
  and some with very long multi-year periods \citep[e.g.,][]{skarka_2014}.
The most famous of these is the Blazhko effect \citet{blazhko1907}, 
 which is a periodic modulation of the amplitude and shape
 of an RRL light curve, with periods of order of tens to hundreds of pulsation cycles.
The amplitude modulation from the Blazhko effect varies from star to star (and by passband) with 
 ranges of a few hundredths to several tenths of a magnitude.

In addition, RRL stars can show period changes thought to correspond
 to the evolutionary path of an individual star. 
Such changes can appear to be sudden or gradual, depending on the time sampling of data sets.
Recent efforts have combined temporally well-sampled, long-baseline photometric data sets 
 in Galactic star clusters to explore these effects and find that most period changes { ascribed to evolutionary effects} are consistent with being gradual when visualized with semicontinuous sampling over very long timescales.
{ Period changes attributed to nonevolutionary origins are also observed, 
 and in contrast to evolutionary effects, these are identified as sudden nonlinear or chaotic evolution of the light curve over time
\citep[for example see][for a century-long study of period changes in M5]{ferro2016}.} 
In our sample, nine stars show the Blazhko effect from \citet[][16\%; their Table 5.2]{smith_1995}, an additional four stars show the Blazhko effect from { the compilation of} \citet[][7\%;]{skarka_2013}, and two stars show the Blazhko effect in \citet[][4\%]{skarka_2014}, for a total of 15 stars showing the Blazhko effect. 
{ References for the studies demonstrating the Blazhko effect are given for each affected star in Appendix \ref{app:indstars}.}

The total frequency of Blazhko stars in our sample, 27\%, is comparable to that observed for the full field RR Lyrae population and in clusters { \citep[e.g.,][though recent work has suggested this number could be as high as 50\% when small amplitude variations are included; J. Jurcsik, private communication]{smith_1995}.}
The Blazhko periods for our sample (see Table \ref{tab:sample}) range from tens of days to several years
 and have Blazhko amplitudes that can be as small as only a few hundredths of a magnitude
  or as large as a few tenths of a magnitude.
{In our sample, 32\% of our RRab stars are Blazhko (12 stars)
 and 18\% of our RRc stars are Blazhko (3 stars); }
these frequencies are similar to but not perfectly matched to the general statistics for the Galactic RRL population, where $\sim$50\% of RRab and $\sim$10\% of RRc type variables show amplitude variations \citep[{ though we note the small number statistics for our RRc stars}; for discussion see Section 6.5 of][and references therein]{catelan_2015}.
Considering our overall breakdown in RRL subtypes and amplitude modulation effects, the demographics 
 of our RRL sample are not unrepresentative of the broader field population of { Galactic} RRL.
 
We note that our discussion has focused on stars with a single pulsation mode. 
In the era of space-based photometry and top-quality photometry gathered from Earth \citep[e.g. \emph{Kepler} from space and OGLE from the ground being two representative examples; see][respectively]{nemec_2013,smolec_2015}, a more complicated view of RRL stars has emerged, and they can no longer be considered `simple' radially pulsating stars.
In 200 day continuous coverage for RRL in the M3 star cluster \citet{jurcsik_2015}, find that 70\% of RRc stars show multi-periodicity.
Moreover, nonradial pulsation in RRc stars appears to be common because 14 out of 15 RRc stars (93\%) observed from space show evidence for additional nonradial modes \citep[e.g.,][]{szabo_2014, moskalik_2015, molnar_2015, kurtz_2016}.
Similarly, the incidence rate in the top-quality ground-based data is also large, with 27\% of RRc in the OGLE Galactic bulge data \citep{netzel_2015} and 38\% in M3 photometry \citep{jurcsik_2015} showing nonradial modes. 
Because many of these effects are of overall small amplitude, these nonradial modes and additional periodicities
 likely have a negligible effect on the goals of this work -- to produce a robust Galactic calibration sample for parallax-based PL determinations -- but do suggest that there is great complexity to the physical mechanisms at work in RRL stars.
With these considerations in mind, an empirical approach to calibrating PL or PLZ relations is indeed preferable to theoretically derived relations that typically do not include the all of the physics driving these nonradial and multimode effects.

\subsection{Effectiveness of Sample for PL Determinations}

Figure \ref{fig:rrldist}(a) shows the sky distribution of the CRRP targets for Galactic coordinates.
The stars span the complete range of R.A. and cover both hemispheres.  The bulk of our sample of Galactic RRL reside out of the Galactic plane,  which reduces complications from large line-of-sight extinction and variations of the reddening law within the disk \citep[e.g.,][among others]{zasowski_2009}.

bf Figure \ref{fig:rrldist}(b) shows the $\text{[Fe/H]}$ distribution of our catalog, spanning 2.5~dex in metallicity with the most metal poor star at $\text{[Fe/H]}= -2.56$ (UY Boo) and the most metal rich at $\text{[Fe/H]}= -0.07$ (AN Ser).
The $\text{[Fe/H]}$ values are adopted from the compilation of \citet{fernley_1998} {\bf (as presented in \citet{feast_2008}}) for all but two stars; 
 the $\text{[Fe/H]}$ for BB Pup comes from \citet{fernley_1998letter} 
 and that for CU Com from \citet{clementini_2000}.
The values from \citet{fernley_1998} come from both direct high-resolution measurements and the $\Delta S$ method of \citet{preston1959}, 
 with the latter being calibrated to the former and brought onto a uniform scale having typical uncertainties of 0.15~dex\footnote{We note for the benefit of the reader that this information on the origin of the [Fe/H] measurement for each star is embedded in the online-only data table associated with the manuscript of \citet{fernley_1998} and not included in the text of that manuscript. { Note 8 in the online notes to the table provides the references for individual stars as well as the metallicity scale and technique.} A direct link to these data is as follows: \url{ http://vizier.cfa.harvard.edu/viz-bin/VizieR?-source=J/A+A/330/515}}.
The majority of the sample can be considered metal poor with 70\% (38 stars) of the sample uniformly distributed over a range of 1~dex ($-2 < \text{[Fe/H]} < -1$) with a mean value of $\langle \text{[Fe/H]} \rangle= -1.50$~dex ($\sigma= 0.24$~dex).  
The other 30\% (17 stars) make up a metal rich sample that is itself uniformly distributed over a range of 1~dex ($-1 < \text{[Fe/H]} < 0$) with a mean value of $\langle \text{[Fe/H]} \rangle= -0.50$~dex ($\sigma= 0.30$~dex).  

Figure \ref{fig:rrldist}(c) shows the period distribution of our sample, with the shortest period being $P = 0.25$~days (DH Peg) and the longest $P = 0.73$~days (HK Pup) ($\log(P) = -0.60$ and $\log(P) = -0.14$, respectively). 
The sample is well designed to uniformly cover the distribution of $\log(P)$ anticipated in RRL populations and provide the greatest leverage to determine RRL relationships with respect to period. 

With the exception of CU Com and BB Pup, all of our stars were included in the \hip{} catalog 
 (albeit with fractional errors larger than 30\%),
 { which means most stars were included in the \emph{Tycho-\gaia{} Astrometric Solution} catalog 
  \citep[TGAS;][]{michalik_2015,lindegren_2016} as part of Gaia DR1 \citet{isgaia,gaiadr1}}\footnote{The first \gaia{} data release occurred on 2016 September 14: \url{http://www.cosmos.esa.int/web/gaia/release}}.
All five RR Lyrae from the \emph{Hubble Space Telescope-Fine Guidance Sensor} parallax program 
 presented in \citet{benedict_2011} are included in our sample. 
Seventeen stars have distances derived previously from the Baade-Wesselink (BW) technique in the
recent compilation of \citet[][and references therein]{muraveva_2015}.
An additional five stars have BW distances from other works (see Table \ref{tab:sample}). 
The last { three columns} of Table \ref{tab:sample} summarize the available distance measurements on a star-by-star basis.

\begin{deluxetable*}{cllccclccc}
\tabletypesize{\tiny}
\tablecolumns{10}
\tablecaption{RRL Galactic Calibrators and Ephemerides. \label{tab:sample}}

\tablewidth{0pt}
\tablehead{
\colhead{Name} & \colhead{$P_{\text{final}}$} & \colhead{HJD$_{\text{max}}$} & \colhead{$\zeta$} & \colhead{RRL Type} &
\colhead{P$_{Bl}$} & \colhead{\text{[Fe/H]}\tablenotemark{a}} & \multicolumn{3}{c}{$\pi$\tablenotemark{b}} \\
\colhead{} & \colhead{(days)} & \colhead{(days)} & \colhead{(days yr$^{-1}$)} & \colhead{} &
\colhead{} & \colhead{} & \colhead{HIP} & \colhead{BW} & \colhead{HST}
}

\startdata
    SW And & 0.4422602   & 2456876.9206 &   1.720e-04 &   RRab &        36.8 &   -0.24  & HIP & 1,2 &      \\ 
    XX And & 0.722757    & 2456750.915  &     \nodata &   RRab &     \nodata &   -1.94  & HIP &     &      \\ 
    WY Ant & 0.5743456   & 2456750.384  &  -1.460e-04 &   RRab &     \nodata &   -1.48  & HIP & 3   &      \\ 
     X Ari & 0.65117288  & 2456750.387  &   -2.40e-04 &   RRab &     \nodata &   -2.43  & HIP & 4,5 &      \\ 
    ST Boo & 0.622286    & 2456750.525  &     \nodata &   RRab &       284.0 &   -1.76  & HIP &     &      \\ 
    UY Boo & 0.65083     & 2456750.522  &     \nodata &   RRab & { 171.8} &   -2.56  & HIP &     &      \\ 
    RR Cet & 0.553029    & 2456750.365  &     \nodata &   RRab &     \nodata &   -1.45  & HIP & 1    &      \\ 
     W Crt & 0.41201459  & 2456750.279  &  -9.400e-05 &   RRab &     \nodata &   -0.54  & HIP & 3   &      \\ 
    UY Cyg & 0.56070478  & 2456750.608  &     \nodata &   RRab &     \nodata &   -0.80  & HIP &     &      \\ 
    XZ Cyg & 0.46659934  & 2456750.550  &     \nodata &   RRab &        57.3 &   -1.44  & HIP &     & HST  \\ 
    DX Del & 0.47261673  & 2456750.248  &     \nodata &   RRab &     \nodata &   -0.39  & HIP & 2,8   &      \\ 
    SU Dra & 0.66042001  & 2456750.580  &     \nodata &   RRab &     \nodata &   -1.80  & HIP & 1   & HST  \\ 
    SW Dra & 0.56966993  & 2456750.400  &     \nodata &   RRab &     \nodata &   -1.12  & HIP & 5   &      \\ 
    RX Eri & 0.58724622  & 2456750.480  &     \nodata &   RRab &     \nodata &   -1.33  & HIP & 1   &      \\ 
    SV Eri & 0.713853    & 2456749.956  &     \nodata &   RRab &     \nodata &   -1.70  & HIP &     &      \\ 
    RR Gem & 0.39729     & 2456750.485  &     \nodata &   RRab &   { 7.2} &   -0.29  & HIP & 1   &      \\ 
    TW Her & 0.399600104 & 2456750.388  &     \nodata &   RRab &     \nodata &   -0.69  & HIP & 5 &      \\ 
    VX Her & 0.45535984  & 2456750.405  &  -2.400e-04 &   RRab &{ 455.37} &   -1.58  & HIP &    &      \\ 
    SV Hya & 0.4785428   & 2456750.377  &     \nodata &   RRab &        63.3 &   -1.50  & HIP &    &      \\ 
     V Ind & 0.4796017   & 2456750.041  &     \nodata &   RRab &     \nodata &   -1.50  & HIP &    &      \\ 
    RR Leo & 0.4523933   & 2456750.630  &     \nodata &   RRab &     \nodata &   -1.60  & HIP & 1  &      \\ 
    TT Lyn & 0.597434355 & 2456750.790  &     \nodata &   RRab &     \nodata &   -1.56  & HIP & 1  &      \\ 
    RR Lyr & 0.5668378   & 2456750.210  &     \nodata &   RRab &        39.8 &   -1.39  & HIP &    & HST  \\ 
    RV Oct & 0.5711625   & 2456750.570  &     \nodata &   RRab &     \nodata &   -1.71  & HIP & 3  &      \\ 
    UV Oct & 0.54258     & 2456750.440  &     \nodata &   RRab &       144.0 &   -1.74  & HIP &    & HST  \\ 
    AV Peg & 0.3903747   & 2456750.518  &     \nodata &   RRab &     \nodata &   -0.08  & HIP & 1  &      \\ 
    BH Peg & 0.640993    & 2456750.794  &     \nodata &   RRab &        39.8 &   -1.22  & HIP &    &      \\ 
    BB Pup & 0.48054884  & 2456750.102  &     \nodata &   RRab &     \nodata &   -0.60\tablenotemark{c}  & &3& \\ 
    HK Pup & 0.7342073   & 2456750.387  &     \nodata &   RRab &     \nodata &   -1.11  & HIP &    &      \\ 
    RU Scl & 0.493355    & 2456750.296  &     \nodata &   RRab &        23.9 &   -1.27  & HIP &    &      \\ 
    AN Ser & 0.52207144  & 2456750.334  &     \nodata &   RRab &     \nodata &   -0.07  & HIP &    &      \\ 
 V0440 Sgr & 0.47747883  & 2456750.706  &     \nodata &   RRab &     \nodata &   -1.40  & HIP & 6  &    \\ 
 V0675 Sgr & 0.6422893   & 2456750.819  &     \nodata &   RRab &     \nodata &   -2.28  & HIP &    &    \\ 
    AB UMa & 0.59958113  & 2456750.6864 &     \nodata &   RRab &     \nodata &   -0.49  & HIP &    &    \\ 
    RV UMa & 0.46806     & 2456750.455  &     \nodata &   RRab &        90.1 &   -1.20  & HIP &    &    \\ 
    TU UMa & 0.5576587   & 2456750.033  &     \nodata &   RRab &     \nodata &   -1.51  & HIP & 1  &    \\ 
    UU Vir & 0.4756089   & 2456750.0557 &  -9.300e-05 &   RRab &     \nodata &   -0.87  & HIP & 1,5 &    \\ 
    AE Boo & 0.31489     & 2456750.435  &     \nodata &    RRc &     \nodata &   -1.39  & HIP &    &    \\ 
    TV Boo & 0.31256107  & 2456750.0962 &     \nodata &    RRc &  { 9.74} &   -2.44  & HIP & 1  &    \\ 
    ST CVn & 0.329045    & 2456750.567  &     \nodata &    RRc &     \nodata &   -1.07  & HIP &    &    \\ 
    UY Cam & 0.2670274   & 2456750.147  &   2.400e-04 &    RRc &     \nodata &   -1.33  & HIP &    &    \\ 
    YZ Cap & 0.2734563   & 2456750.400  &     \nodata &    RRc &     \nodata &   -1.06  & HIP & 6  &    \\ 
    RZ Cep & 0.30868     & 2456755.135  &  -1.420e-03 &    RRc &     \nodata &   -1.77  & HIP &    & HST \\ 
    RV CrB & 0.33168     & 2456750.524  &     \nodata &    RRc &     \nodata &   -1.69  & HIP &    &   \\ 
    CS Eri & 0.311331    & 2456750.380  &     \nodata &    RRc &     \nodata &   -1.41  & HIP &    &   \\ 
    BX Leo & 0.362755    & 2456750.782  &     \nodata &    RRc &     \nodata &   -1.28  & HIP &    &   \\ 
    DH Peg & 0.25551053  & 2456553.0695 &     \nodata &    RRc &     \nodata &   -0.92  & HIP & 5,7  &   \\ 
    RU Psc & 0.390365    & 2456750.335  &     \nodata &    RRc &        28.8 &   -1.75  & HIP &    &   \\ 
    SV Scl & 0.377356    & 2457000.479  &     3.0e-04 &    RRc &     \nodata &   -1.77  & HIP &    &   \\ 
    AP Ser & 0.34083     & 2456750.510  &     \nodata &    RRc &     \nodata &   -1.58  & HIP &    &   \\ 
     T Sex & 0.3246846   & 2456750.229  &   1.871e-03 &    RRc &     \nodata &   -1.34  & HIP & 1  &   \\ 
    MT Tel & 0.3168974   & 2456750.108  &   5.940e-04 &    RRc &     \nodata &   -1.85  & HIP &    &   \\ 
    AM Tuc & 0.4058016   & 2456750.392  &     \nodata &    RRc &      1748.9 &   -1.49  & HIP &    &   \\ 
    SX UMa & 0.3071178   & 2456750.347  &     \nodata &    RRc &     \nodata &   -1.81  & HIP &    &   \\ 
    CU Com & 0.4057605   & 2456750.410  &     \nodata &    RRd &     \nodata &   -2.38\tablenotemark{d}& &    &     \\ 
\enddata

\tablenotetext{a}{{ Unless otherwise noted, values are taken from \citet{feast_2008}, but the measurements were first compiled by \citet[][and references therein]{fernley_1998} and are on a metallicity scale defined by \citet[][and references therein]{fernley_1997}.}}
\tablenotetext{b}{Indicates if the star has a parallax derived from \hip~\citep[HIP;][]{hip}, \hst~\citep[HST;][]{benedict_2011}, and/or Baade-Wesselink (BW; source indicated in footnotes).  }
\tablenotetext{c}{\text{[Fe/H]} comes from \citet{fernley_1998letter}.}
\tablenotetext{d}{\text{[Fe/H]} comes from \citet{clementini_2000}.}
\tablerefs{
(1)~\citet{liu_1990}; (2)~\citet{jones_1992}; (3)~\citet{skillen_1993jhk}; (4)~\citet{fernley_1989}; (5)~\citet{jones_1988}; (6)~\citet{cacciari_1989c}; (7)~\citet{fernley_1990}; (8)~\citet{skillen_1989}
}

\end{deluxetable*}


\section{The Three-hundred Millimeter Telescope}\label{sec:tmmt}

There are significant complications of a technical and practical nature
when contemplating the use of traditional telescope/detector systems to study Galactic RRL.
RRL have typical periods of order a day or less (requiring high-cadence, short-term sampling),
 but they can also show significant variations and period changes over time, 
 requiring additional long-term sampling.
These general considerations aside, the 55 Galactic RRL in our sample are noteworthy for being among the brightest RRL in the sky and are therefore too bright for even some of the ``smallest" (i.e., 1 m class) telescopes at most modern observatories. 
Moreover, bright Galactic RRL are relatively rare and they are distributed over the entire night sky, 
 which means that to build a uniform and relatively complete sample of such targets requires a dual-hemisphere effort. 
In direct response to this need, we assembled a devoted robotic 300mm telescope system to obtain modern, high-cadence, optical light curves for these important targets.
In the sections to follow, we describe the telescope system (Section \ref{ssec:hardware}), 
 the observing procedures (Section \ref{ssec:tmmt_obs}), and the resulting
 photometry (Section \ref{ssec:tmmt_phot}),
 with a summary given in Section \ref{ssec:tmmt_sum}.
 
\subsection{TMMT Hardware} \label{ssec:hardware}
The Three-hundred MilliMeter Telescope (TMMT) is a 300mm $f/7.9$ flat-field Ritchey-Chr\'etien telescope\footnote{Takahashi FRC-300. \url{http://www.takahashiamerica.com}} on an AP1600\footnote{Astro-Physics, Inc. \url{http://www.astro-physics.com} \label{apcc}} mount. 
The imager consists of an Apogee\footnote{Andor Technology plc. \url{http://www.andor.com} \label{api} } D09 camera assembly, which includes an E2V42-40 CCD with mid-band coatings and an external Apogee\textsuperscript{\ref{api}} nine-position filter wheel containing U, B, V, R$_c$ \& I$_c$ Bessel filters, a 7nm wide H$\alpha$ filter, and an aluminum blank acting as a dark slide. 
The imaging equipment is connected to the telescope via a Finger Lakes Instruments ATLAS focuser\footnote{Finger Lakes Instrumentation. \url{http://www.flicamera.com}} with adapters custom-made by Precise Parts\footnote{Precise Parts. \url{http://www.preciseparts.com}}. 
The short back focal length of the system and the depth of the CCD in the camera housing 
 precluded the option of using an off-axis guider.  
For the program described here, individual exposures were short enough that guiding was not required.   
The mount contains absolute encoders, which are able to virtually eliminate periodic error, 
  and the pointing model program applies differential tracking rates that correct for 
  polar misalignment, flexure, and atmospheric refraction.  

The system was first tested in the Northern Hemisphere at the 
 Carnegie Observatories Headquarters (in downtown Pasadena, CA) from 2013 August to 2014 August. It was 
 later (2014 September) shipped to Chile and permanently mounted in the Southern Hemisphere in a dedicated building with a remotely controlled, roll-off roof at Las Campanas Observatory (LCO). RRL monitoring observations continued until 2015 July.

The TMMT is controlled by a PC that can be accessed 
 remotely through a VNC or a remote desktop sharing application.
The unique testing and deployment of the TMMT has enabled true full-sky coverage for our RRL campaign, not just keeping the CCD setup consistent but also using the same system in both hemispheres so to minimize the effects of observational (equipment) systematics on our science goals.  

\subsection{Observations} \label{ssec:tmmt_obs}

ACP Observatory Control Software\footnote{ACP is a trademark of DC-3 Dreams. \url{http://acpx.dc3.com}} is used to automate the actions of the individual hardware components and associated software programs; more specifically,
MaximDL \footnote{MaximDL by Diffraction Limited.  \url{http://www.cyanogen.com/}} controls the camera, 
FocusMax\footnote{FocusMax. \url{http://www.focusmax.org}} controls the focus, and APCC\textsuperscript{\ref{apcc}} controls the mount.  
Weather safety information was obtained via the internet from the nearby HAT South facility 
\footnote{For information see \url{http://hatsouth.org/}}.

For the RRL program, an ACP script was automatically generated each day to observe RRL program stars at phases that had not yet been covered. 
The script included observations of standard stars spaced throughout the night to calibrate the data.  
Additional scripts control other functions of the telescope, including: 
 automatically starting the telescope at dusk, monitoring the weather, 
 monitoring the state of each device and software to catch errors (and restart if necessary),
 and, finally, to shutting down at dawn.      
Images were automatically pipeline-processed using standard procedures, 
 which are detailed in Appendix \ref{app:img_prox}.

The goal of the program was to obtain complete phase coverage for each of the 55 sources.
{ Due to time and observability constraints, in particular the limited time available for the Northern sample while the telescope was deployed at the Carnegie Observatory Headquarters, there are still phase gaps for most of the stars.} 
Additional phase sampling was obtained for nearly completed stars only when it was observationally efficient to do so. 
At the conclusion of the TMMT program, we have photometric observations 
 in the $B$, $V$, and $I_C$ broadband filters for each of our 55 RRL.

\begin{deluxetable}{lccccccccc}
\tablewidth{0pt}
\tabletypesize{\tiny}
\tablecaption{Optical Photometric Parameters for SBS and LCO. \label{tab:optcalib}}
\tablehead{
\colhead{Site}           & \colhead{$v_1$\tablenotemark{a}}      &
\colhead{$v_2$}          & \colhead{$v_3$}  &
\colhead{$b_1$}          & \colhead{$b_2$}    &
\colhead{$b_3$}  & \colhead{$i_1$}  &
\colhead{$i_2$} & \colhead{$i_3$} 
}
\startdata
SBS                &  -4.18  &  0.23  &   -0.07  &  -0.44  &   0.14   &  1.32  &   0.78  &  0.08  &  0.92  \\
$\sigma_{SBS}$     &   0.10  &  0.10  &    0.01  &   0.15  &   0.15   &  0.03  &   0.13  &  0.09  &  0.02 \\
LCO                &  -4.09  &  0.13  &   -0.08  &  -0.42  &   0.10   &  1.32  &   0.68  &  0.08  &  0.92  \\ 
$\sigma_{LCO}$     &   0.06  &  0.04  &    0.01  &   0.06  &   0.05   &  0.02  &   0.07  &  0.05  &  0.01  \\
\enddata
\tablenotetext{a}{The zero-point is relative to 25, which is the default instrumental magnitude zero-point in \tt{DAOPHOT}.  }
\end{deluxetable}

\subsection{Photometry} \label{ssec:tmmt_phot}

Instrumental magnitudes ($m_B$, $m_V$, $m_{I_C}$) were extracted 
 from the TMMT imaging data using \texttt{DAOPHOT} \citep{stetson_1987}.
Aperture corrections were measured using \texttt{DAOGROW} \citep{stetson_1990} and applied to correct the aperture photometry to infinite radius. 

The magnitudes ($B$, $V$, $I_C$) and colors ($B-V$) and ($V-I_C$) for a set of standard stars were adopted from Landolt standard fields \citep{cousins_1980,landolt_1983,cousins_1984,landolt_2009}. 
Photometric calibrations were determined using the \texttt{IRAF} \texttt{PHOTCAL} package \citep{1993ASPC...52..479D} with the \texttt{fitparams} and \texttt{invertfit} tasks. 
Second-order extinction terms were also measured, but found to be negligible and therefore not included. 

The final adopted photometric calibration procedure is as follows:
The zero-points are referenced to an airmass of 1.5 to minimize correlation between airmass ($X_{B}$, $X_{V}$, $X_{I_C}$) and zero-point terms. 
For the purposes of this work, we provide our calibrations for three of the photometric bands, $m_B$, $m_V$, $m_{I_C}$. 
We define relationships between these instrumental magnitudes and the true magnitudes ($B$, $V$, $I_C$) and  colors ($B-V$, $V-I_C$) as follows: 
\begin{equation}
V = v_1 + m_V-v_2 \left[(X_V)-1.5\right]+v_3 (m_V-m_{I_C})
\end{equation}
\begin{equation}
B-V = b_1 - b_2 \left[0.5 \times (X_B + X_V) -1.5\right] + b_3(m_B-m_V)
\end{equation}
and 
\begin{equation}
 V-I_C = i_1 - i_2 \left[0.5 \times (X_{I_C} + X_V) - 1.5\right] + i_3(m_V-m_{I_C})
\end{equation}
The coefficients for the various corrections are:
 (i) zero-point offset ($b_1$, $v_1$, $i_1$),
 (ii) airmass term ($b_2$, $v_2$, $i_2$),
 (iii) and a color term ($b_3$, $v_3$, $i_3$).
These three sets of coefficients (one set for each filter) are unique for the telescope and observing site.
Median values and 1$\sigma$ deviations for the TMMT at the Carnegie Observatories Santa Barbara Street (SBS) and Las Campanas Observatory (LCO) sites
 are given in Table \ref{tab:optcalib}, 
 and were used as the starting point for calibrating individual nights or 
 were adopted as the solution if not enough standards were observed.   
Each photometric night has a systematic zero-point error determined from fitting the standard star photometry, which propagates to each RRL observed on that night. 
To reduce the final systematic uncertainty on the mean magnitude of an RRL, each RRL was observed on as many photometric nights as possible.  

\begin{figure}
\includegraphics[width=\columnwidth]{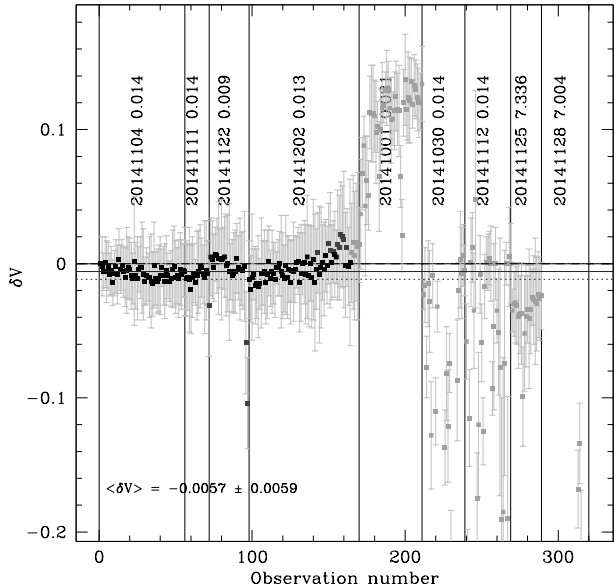}
\caption{\label{fig:stdmag}
An example of the photometric errors over multiple nights for V Ind.  
A total of 326 images were reduced over nine nonconsecutive nights.  Each frame is calibrated based on the transformation equations for that night. The average difference magnitude ($\delta V$) for stars relative to the first frame is plotted in with associated error bars. 
Four of the nights (black points) were photometric and had good photometric solutions from observations of standard star.
The remaining nights either had poor solutions or used the default transformation coefficients. 
The systematic error of the nightly zero-point solution is given next to the date of observation; see text for details.  
}
\end{figure}

\texttt{DAOMATCH} and \texttt{DAOMASTER} 
 were used to create the light curve relative to the first frame by finding and subtracting the average magnitude offset (determined from the ensemble photometry of common stars in  all the frames) relative to the first frame in the series.  
 Figure \ref{fig:stdmag} shows the zero-point data for V Ind. The average differential magnitude ($\delta V$) for stars relative to the first frame is plotted with associated error bars.
Since each frame was calibrated any frame taken on a photometric night could act as the reference, or alternatively, the average of all the photometrically calibrated frames can be used. The advantage of using the average is that it minimizes random fluctuations or poorly calibrated frames. Figure \ref{fig:stdmag} illustrates an example where the first frame calibration deviated only slightly from the average.  
The final photometry is corrected by the average offset relative to the reference frame, thus avoiding the problem of choosing the ``best'' frame.  
Since multiple nights were used, each independently calibrated, the final systematic error is reduced.  
 Note that the two nights plotted on the right of Figure \ref{fig:stdmag} were not photometric and the default photometric solution was adopted. The transformation errors on nights such as this may appear discrepant if too few stars were used, resulting in potentially unrealistic photometric solutions; hence these nights are not used for calibration. Trends in nonphotometric data may be correlated with airmass, in which case the airmass term may be poorly constrained.  Dips in the data may be due to a passing cloud or variable conditions. 

\subsection{Summary} \label{ssec:tmmt_sum}

The TMMT is a fully robotic, 300 mm telescope at LCO, 
 for which the nightly operation and data processing have completely automated. 
Over the course of two years data were collected on 179 individual nights for
 our sample of the 55 RR Lyrae in the $B$,$V$, and $I_C$ broadband filters. 
Of these nights, 76 were under photometric conditions and calibrated directly.
The 103 nonphotometric nights were roughly calibrated by using the default transformation equations, but only provide differential photometry relative to the calibrated frames.  
This resulted in 59,698 final individual observations.  
Individual data points have a typical photometric precision of 0.02 mag.  
The statistical error falls rapidly with hundreds of observations, with the zero-point uncertainties being the largest source of uncertainty in the final reported mean magnitude.  
%
\section{Archival Observations} \label{sec:data}
 RR Lyrae variables can show changes in their periods \citep[see discussions in][]{smith_1995}, and can have large accumulated effects from period inaccuracies,
  making it problematic to apply ephemerides derived from earlier work to new observational campaigns.
While these two causes--- period changes and period inaccuracies--- have very different
 physical meanings (one intrinsic to the star and one to limited observations) the
 effect on trying to use data over long baselines is the same:
  individual observations will not ``phase up'' to form a self-consistent light curve.
When well-sampled observations are available that cover a few pulsation cycles, it is possible to visually see the phase offset and simply align light-curve substructure (i.e., { the exact timings of minimum and maximum light}), but in the case of sparse sampling the resulting phased data will not necessarily form clear identifiable sequences (more details will be given in Section \ref{sec:datamerge}).

 Thus, to compare the results of our TMMT campaign to previous studies of these RRL and to fill gaps in our TMMT phase coverage,
  we have compiled available broadband data from literature published over the past 30 years and spanning our full wavelength coverage. We note that this is not a comprehensive search of all available photometry.
 In the following sections, we give an overview of data sources for the sample, organized by passband,
  with star-by-star details given in Appendix \ref{app:indstars}.
 Optical observations are described in Section \ref{sec:opticaldata},
  NIR observations in Section \ref{sec:nirdata},
  and MIR observations in Section \ref{sec:mirdata}.
 Observations are converted from their native photometric systems
  to Johnson $U$, $B$, $V$, Kron-Cousins $R_C$,$ I_C$, 2MASS \JHK, and \emph{Spitzer} 3.6$\mu$m and 4.5$\mu$m,
   with the transformations given in the text. 
 Section \ref{sec:datasummary} presents a summary of the resulting archival datasets. 

\subsection{Optical Data} \label{sec:opticaldata}
\subsubsection{ASAS} \label{sec:asas}
{ The All Sky Automated Survey}\footnote{\url{http://www.astrouw.edu.pl/asas/}} (ASAS) 
 is a long-term project monitoring all stars brighter than $V\sim$14 mag 
 \citep{pojmanski_1997,Pojmanski_2002,Pojmanski_2003,Pojmanski_2004,Pojmanski_2005a,Pojmanski_2005b}. 
The program covers both hemispheres, with telescopes at Las Campanas Observatory in Chile 
 and Haleakala on Maui, both of which provide simultaneous $I$ and $V$ photometry. 
Not all photometry produced by the program has yet been made public
 (i.e., only $I$ or $V$ is available and for only limited fields and time frames). 
Moreover, several of our brightest targets, 
 for example SU Dra and RZ Cep, both of which have parallaxes from the \emph{HST-FGS} program, are not included.
We adopt $V$ magnitudes from ASAS to augment phase coverage for some of our sample, if needed and where available.

\subsubsection{GEOS} \label{sec:geos}
The { Groupe Europ$e^{'}$en d'Observations Stellaires} (GEOS) { RR Lyr Survey} \footnote{\url{http://www.ast.obs-mip.fr/users/leborgne/dbRR/grrs.html}} 
 is a long-term program utilizing TAROT\footnote{\url{http://tarot.obs-hp.fr/}} \citep{klotz_2008,klotz_2009}
 at Calern Observatory (Observatoire de la Côte d'Azur, Nice University, France).
Annual data releases from this project add times for { maximum light} for program stars over
 the last year of observations
 \citep[data releases include][among others]{geos2005,geos2006,geos2006b,geos2007,geos2007b,geos2008,geos2009,geos2011,leborgne_2013}. 
GEOS aims to characterize period variations in RRL stars by
 providing long-term, homogeneous monitoring of bright RRL stars,
  albeit only around the anticipated times of maximum light.
The primary public data product from this program are the times of light curve maxima over a continuous period
 since the inception of the program in 2000.
A well observed star will have its maximum identified to a precision of 4.3 minutes (0.003 days),
 but measurements vary between $\sigma_{max}$=0.002 and $\sigma_{max}$=0.010 days depending on local weather conditions.
Such data are invaluable for understanding period and amplitude modulations for specific RRL
 \citep[e.g., RR Lyr in][]{leborgne_2014} and for RRL as a population \citep{leborgne_2007,leborgne_2012}.
We utilize the timing of maxima provided by GEOS for our common stars, primarily for the phasing efforts to be described in Section \ref{sec:datamerge}. 
 
\subsubsection{Individual Studies}
In addition to the large programs previously described,
 we use data from individual studies over the past 30 years. 
Due to the diversity of such works, 
 we must determine filter transformations on a study-by-study basis as we now describe.

For the conversion of Johnson $R_J$ and $I_J$ to Cousins $R_C$ and $I_C$
 (noting that $V$ is the same in either),
 we utilize the following transformations from \citet{fernie_1983}:
\[ 
V - R_C = 
\begin{cases}
 -0.024 + 0.730\times(V-R_J)  & \text{for }  (V-R_J)\leq1.1  \\
 +0.218 + 0.522\times(V-R_J)  & \text{for }  (V-R_J) >1.1 
\end{cases}
\]
and 
\[ 
R_C - I_C = 
\begin{cases}
 +0.034 + 0.845\times(R_J-I_J)  & \text{for } (R_J-I_J)\leq0.8  \\
 -0.239 + 1.315\times(R_J-I_J)  & \text{for } (R_J-I_J)>0.8 
\end{cases}
\]
\[ 
V - I_C = 
\begin{cases}
 +0.004 + 0.783\times(V-I_J)  & \text{for } (V-I_J)\leq1.9  \\
 -0.507 + 1.217\times(V-I_J)  & \text{for } (V-I_J)>1.9 
\end{cases}
\]
This is required to convert data from \citet{Barnes_1992} and \citet{cacciari_1987}
 to our system.

Optical data from \citet{skillen_1993,liu_1989,Pac_1965_2, jones_1992,Warren_1966,fernley_1990,fernley_1989}
 and \citet{clementini_2000} are also used for our study but no global transformations were required.

\subsection{Near-Infrared Data} \label{sec:nirdata}

Multi-Epoch data in the NIR are particularly sparse, but owing to numerous RRL campaigns in the 
  1980s and 1990s to apply the BW technique to determine distances to these stars,
  there are some archival data in these bands.
 Care, however, must be taken in using these archival data directly with more recent data, because (i)
  they must be brought onto the same photometric system (filter systems and detector technology have changed)
  and (ii) RRL are prone to period shifts over rather short time-scales.
 While the former concern can be characterized statistically, 
  the latter concern presents a serious limitation to the use of archival data. 
 Contemporaneous optical observations are necessary to properly phase the NIR data with our modern optical data.
 Thus, only data that could be phased, owing to the availability of contemporaneous optical data,
  could ultimately be used for our purposes. 

\subsubsection{2MASS} \label{sec:2massdata}
Single-epoch photometry is available from 2MASS \citep{skrutskie_2006} in $J$, $H$, and $K_s$. 
Phasing of these data was accomplished primarily using data from GEOS (Section \ref{sec:geos}).

\subsubsection{Individual Studies} \label{sec:nirind}

Data from \citet{sollima_2008} are adopted and already on the 2MASS system.

To convert from the CIT systems to 2MASS we use the following transformations:\footnote{\url{http://www.astro.caltech.edu/~jmc/2mass/v3/transformations/}}
\begin{multline}
(J-K)_{\text{2MASS}} = 1.068\times(J-K)_{\text{CIT}}-0.020 \\
(H-K)_{\text{2MASS}} = 1.000\times(H-K)_{\text{CIT}}+0.034 \\ 
K_s = K_{\text{CIT}} - 0.019 + 0.001\times(J-K)_{\text{CIT}} 
\end{multline}
This was required for data presented in \citet{liu_1989}, \citet{Barnes_1992} and \citet{fernley_1993}.
If no color was provided, then the average color for RRL of $(J-K)_{\text{CIT}}$ = 0.25 was adopted.

The data in \citet{skillen_1989} and \citet{fernley_1990} required conversion from the UKIRT system as follows:
\begin{multline}
(J-K)_{\text{2MASS}} = 1.070\times(J-K)_{\text{UKIRT}} - 0.015 \\
(H-K)_{\text{2MASS}} = 1.071\times(H-K)_{\text{UKIRT}} + 0.014 \\
K_s = K_{\text{UKIRT}} + 0.003 +0.04\times(J-K)_{\text{UKIRT}}
\end{multline}
The average colors for RRL of $(J-K)_{\text{UKIRT}}$ = 0.3
 and $(H-K)_{\text{UKIRT}}$ = 0.1 were adopted if no color information data were available. 

Data from \citet{fernley_1989} and \citet{skillen_1993jhk} required conversion from the SAAO system to 2MASS as follows:
\begin{multline}
(J-K)_{\text{2MASS}} = 0.944\times(J-K)_{\text{SAAO}} - 0.005 \\
(H-K)_{\text{2MASS}} = 0.945\times(H-K)_{\text{SAAO}} + 0.043 \\
K_s = K_{\text{SAAO}} -0.024 +0.017\times(J-K)_{\text{SAAO}}
\end{multline}
The average colors for RRL of $(J-K)_{\text{SAAO}}$ = 0.2
 and $(H-K)_{\text{SAAO}}$ = 0.2 were adopted if no color data were available. 


\subsection{Mid-Infrared Data} \label{sec:mirdata}

\subsubsection{Spitzer}\label{sec:spitzerdata}
The mid-infrared [3.6] and [4.5] (hereafter also S1 and S2, respectively)  observations were taken using {\it Spitzer}/IRAC as part of the Warm--{\it Spitzer} Exploration Science Carnegie RR Lyrae Program \citep[CRRP; PID 90002][]{spitzer_2012}. 
Each star was observed a minimum of 24 times (with additional observations provided by the {\it Spitzer} Science Center to fill small gaps in the telescope's schedule). 
{The {\it Spitzer} images were processed using SSC pipeline version S19.2. Aperture photometry was performed using the SSC-contributed software tool \texttt{irac\_aphot\_corr}\footnote{\url{http://irsa.ipac.caltech.edu/data/SPITZER/docs/dataanalysistools/tools/contributed/irac/iracaphotcorr/}}, which performs the pixel-phase and location-dependent corrections. The photometry was calibrated to the standard system defined by \citet{reach_2005} using the aperture corrections for the S19.2 pipeline data provided to us by the SSC (S. Carey, 2016, private communication).}

\subsubsection{WISE} \label{sec:wisedata}
\emph{WISE} \citep{wise_survey_paper} 
 or \emph{NEOWISE} \citep{neowise_survey_paper, neowise_survey_paper2}
 photometry is available for each of our stars \citep{wise_survey_paper}. 
This is the only MIR data for three RRL. 
We opt to tie the \emph{WISE} photometric system to that defined for \spitzer. 
For our RRL stars, we find an average offset of 
\begin{equation} \label{eq:wise1_transform}
  W1-[3.6] = -0.038\pm0.010 
\end{equation}
and
\begin{equation} \label{eq:wise2_transform}
  W2-[4.5] = -0.027\pm0.010
\end{equation}
This offset is applied to all of the {\it WISE} or {\it NEOWISE} data used in this work. 

\subsection{Summary of Archival Data} \label{sec:datasummary}

 We have compiled a heterogeneous sample of data in order to build well sampled light curves from the optical to mid-infrared. 
 We have homogenized these diverse data sets to the following filter systems: 
  Johnson $UBV$, Kron-Cousins $RI$, 2MASS \JHK, 
   and  \spitzer~$[3.6]$, $[4.5]$. In addition to the photometric measurements, the phasing of the archival data had to be aligned to the current epoch because the periods of RRL can change over time; see Fig Figure 3 for examples of phase drift.  Aligning the archival data in phase is the topic of Section 5.

Representative light curves for a subset of stars are shown in Figure \ref{fig:examples}.
The stars shown in Figure \ref{fig:examples} were selected based on having good sampling 
 for the bulk of the 10 photometric bands.
Individual data points are given in the first phase cycle of the data ($0 < \phi < 1$),
 with filled symbols being new data from this work (TMMT and \spitzer)
 and open symbols representing data taken from the literature.  
Single-epoch 2MASS data are represented by open pentagons. 
NIR data from \cite{fernley_1993} are available for many stars but only for a few epochs; these data are represented by open triangles.  
All other literature data in the NIR are represented by open squares.   

{\bf Visualizations of this type are provided for each of our 55 stars as a figure set associated with Figure \ref{fig:examples}.}
\begin{figure*}
\begin{center}
 \includegraphics[width=0.45\textwidth]{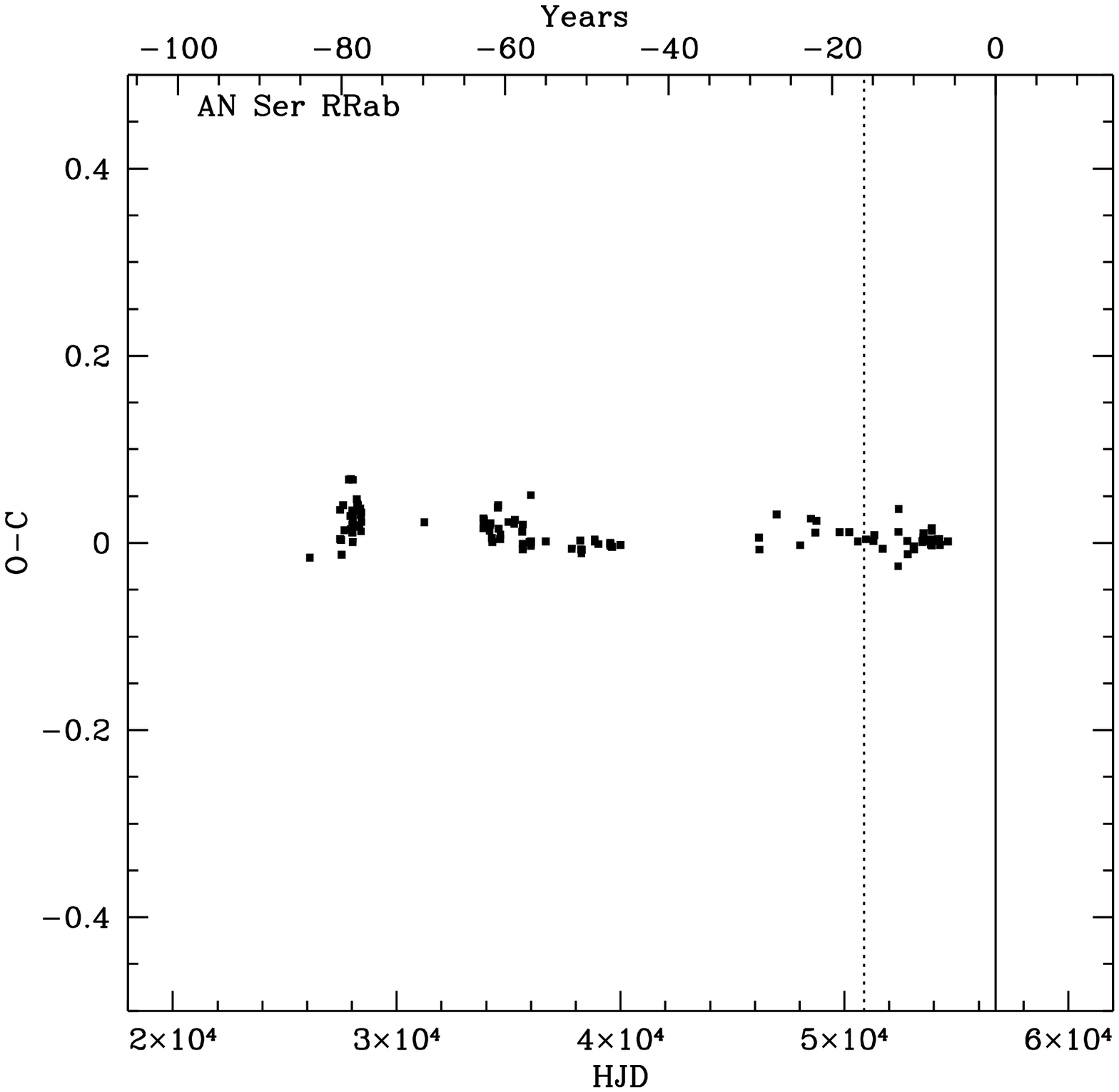} 
 \includegraphics[width=0.45\textwidth]{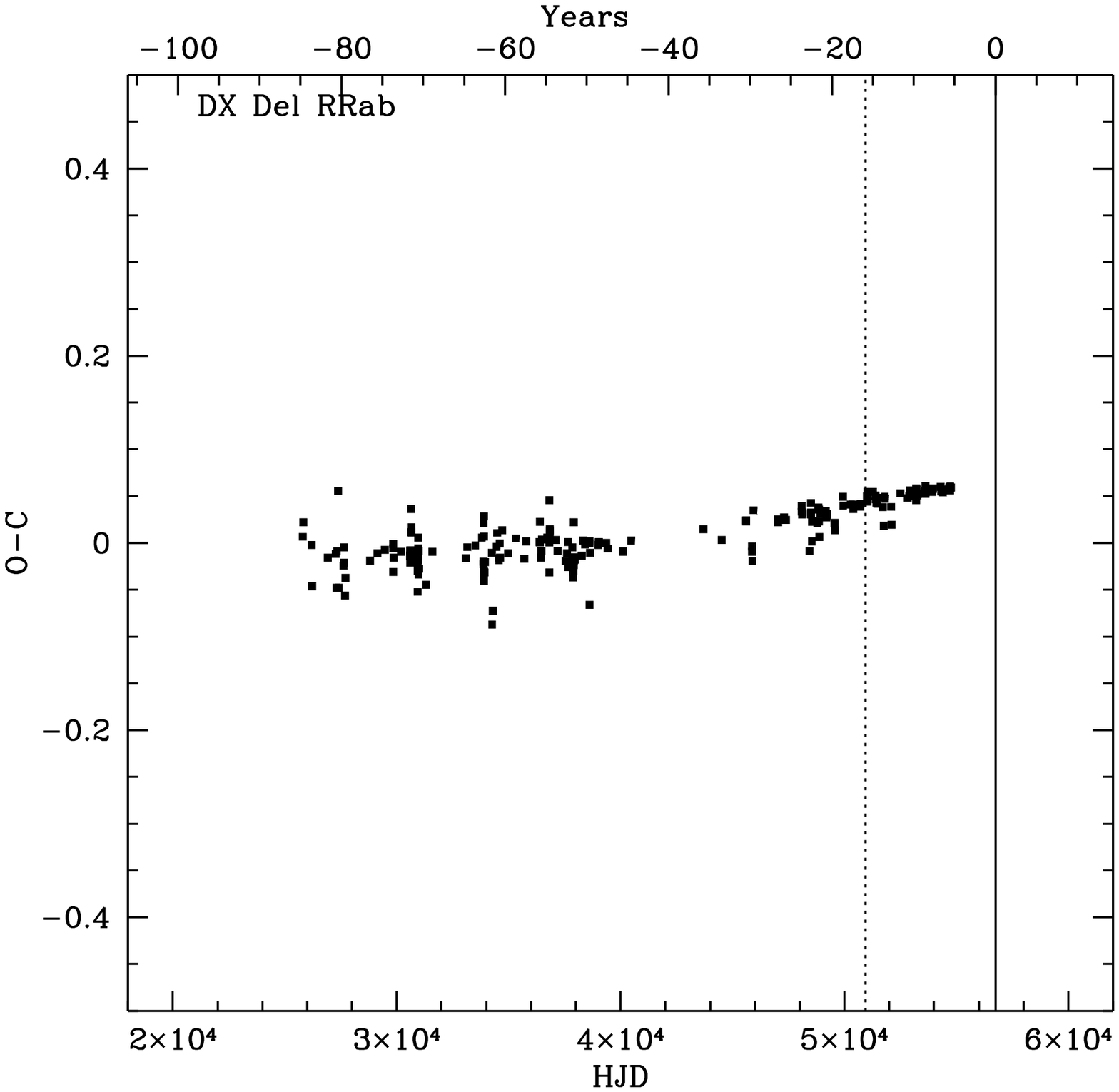}
 \includegraphics[width=0.45\textwidth]{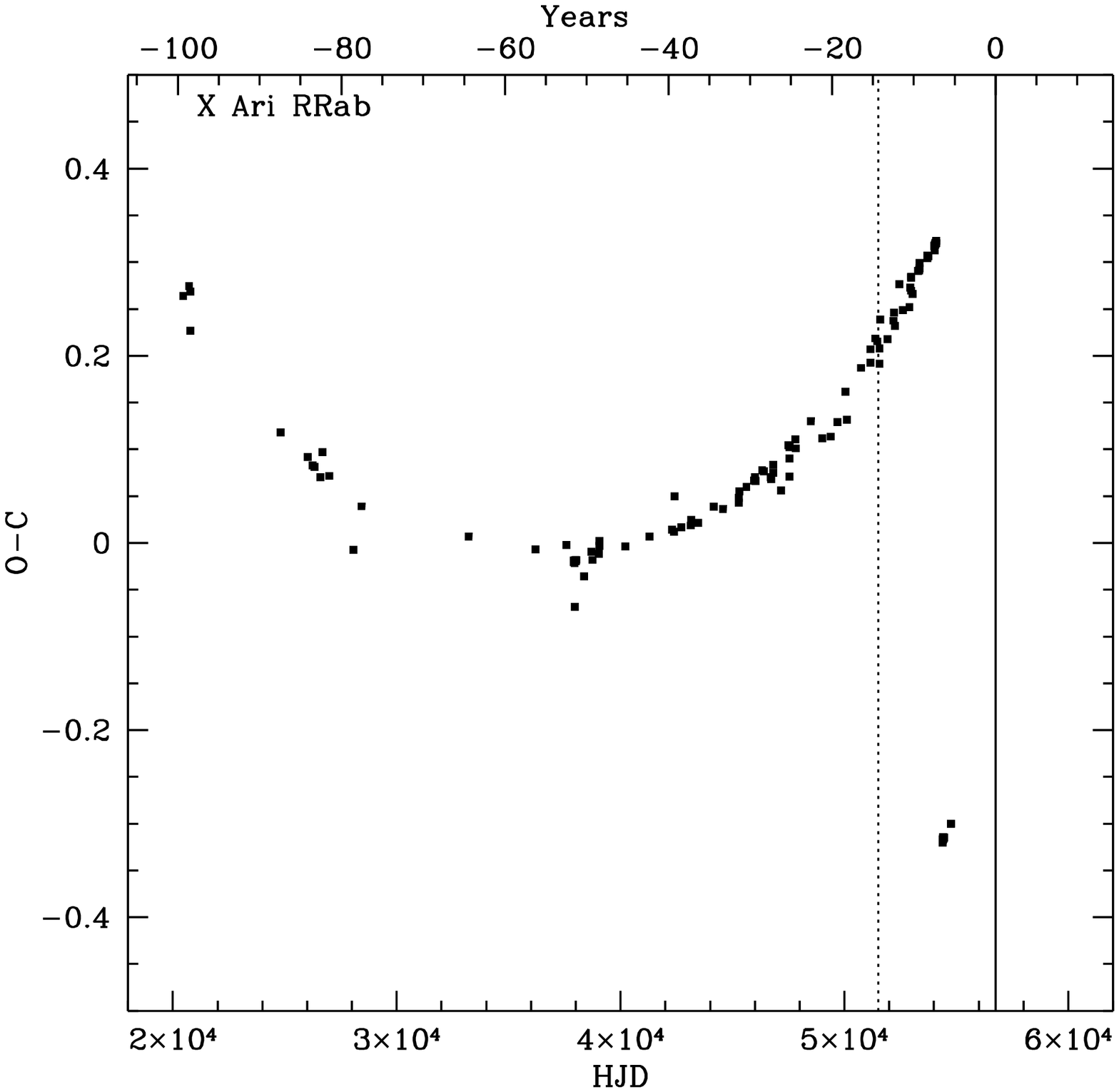}
 \includegraphics[width=0.45\textwidth]{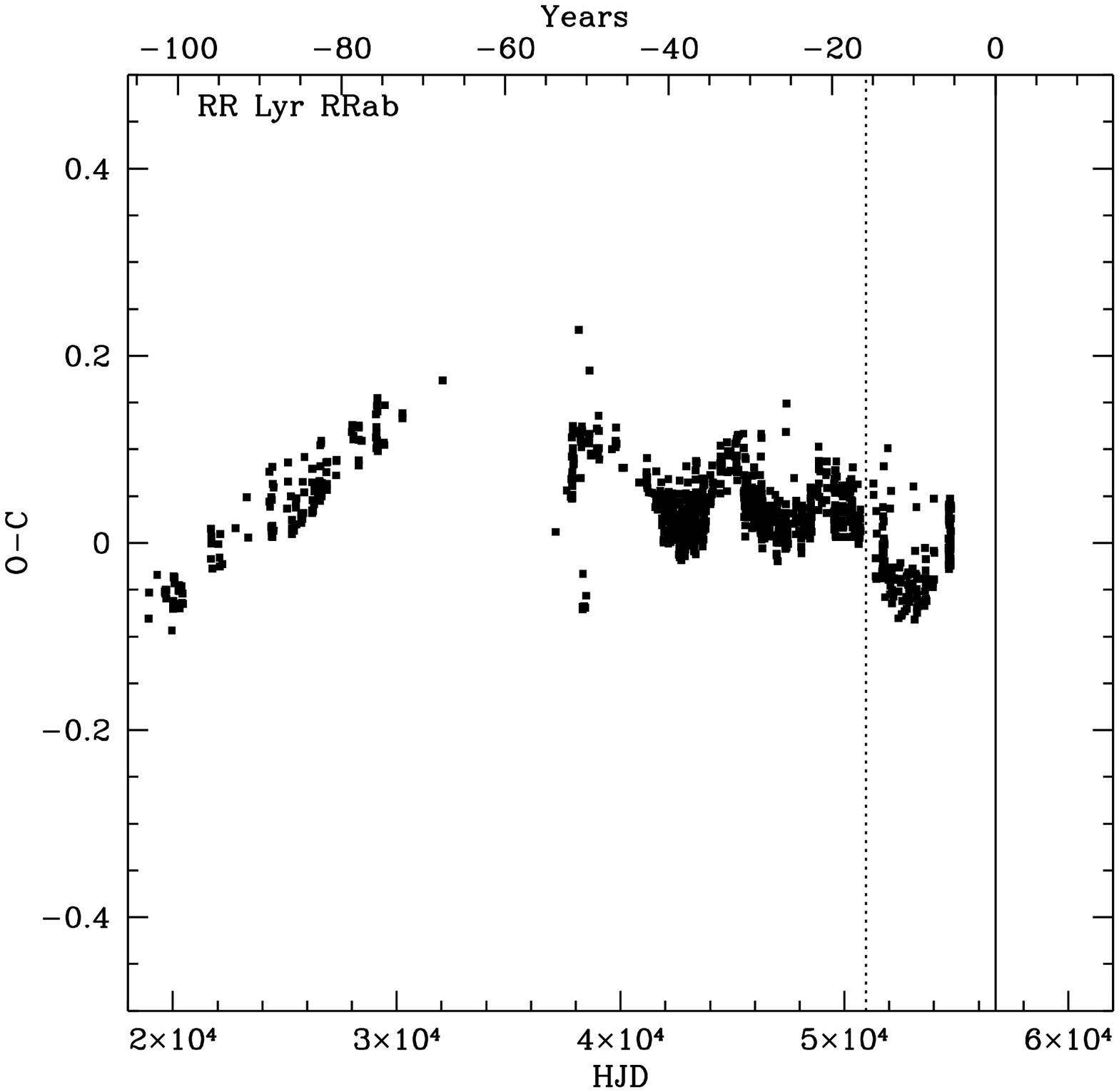}
\caption{ \label{fig:phase}
Data from GEOS (Sec \ref{sec:geos}) demonstrating of phasing solutions in the O-C diagram {\bf (see Section \ref{ssec:ocdiagram} for details)} for the four cases described in the text.  The solid vertical line highlights the epoch of the TMMT data and the vertical dashed line is at the epoch of the 2mass observations.  
 {\bf (a) -- top left} Flat O-C behavior (for AN Ser) implies that no updates to period are required.
 {\bf (b) -- top right} Linear O-C behavior over the last 30 years (for DX Del) implies that the period is slightly longer than originally determined.  
 {\bf (c) -- bottom left} Quadratic O-C behavior (for X Ari) implies that the period itself has been changing over time. 
 {\bf (d) -- bottom right} Chaotic O-C behavior (for RR Lyr) implies that neither adding precision to the period nor applying a shift in period is sufficient to phase all of the individual data-sets analytically. In this case, each data-set gets a custom phase offset for alignment to the current TMMT epoch.  
 } 
\end{center}
\end{figure*}
\section{Reconciling Phase} \label{sec:datamerge}
 
Due to their short periods, RRL experience hundreds to over one thousand period cycles over a single year 
 (between 500 and 1460 cycles per annum for our longest- and shortest-period RRLs, HK Pup and DH Peg, respectively). 
On these long timescales RR Lyrae can show physical changes in their 
 periods due to their own stellar evolution that may appear gradual/smooth or sudden { or they can
  show sudden changes with unexplained origins (i.e., changes not based on evolution)}.
Over our 30 year baseline, a period uncertainty of 1 s results in maximal offsets of 0.17 day or a 0.24 phase for HK Pup (longest period)
 and 0.50 day or 1.95 full cycles for DH Peg (shortest period).
 Each individual observation taken within this time span would have its own phase offset.
In this section, we describe our procedures to merge the data described in the previous section
 by updating the ephemerides for each star. 

For clarification, we define several terms for the discussions to follow.
A data set is either a single or a set of individual photometric measurement(s) for a given star 
 and the associated date of observation taken with the same instrument setup 
 and transformed into our standard photometric systems. 
We define the {\tt HJD$_{\text{max}}$} as the {\tt HJD} at maximum light
 measured from our TMMT data,\footnote{This involves using the \gloess{} light curves for those data that are not fully sampled at maximum. The process for making these light curves is described in Section \ref{sec:gloess}.}
 which means that the quantity is defined uniformly for all stars in our sample.
Our term ($\vartheta$) is the offset in days between our {\tt HJD$_{\text{max}}$} and the date of observation, 
 which is defined mathematically as follows:
\begin{equation} \label{eq:cycles}
 \vartheta_{i} = \text{HJD}_{i,\text{obs}} - \text{HJD}_{\text{max}}
\end{equation}
for each individual data point ($i$). 
Then, we define the initial phase ($\phi^{0}$) for each data point ($i$) as follows:
\begin{equation} \label{eq:initialphase}
\phi^{0}_{i} = \frac{\vartheta_{i}}{P_{\text{archival}}} - \rm{int}{\left( \frac{\vartheta_{i}}{P_{\text{archival}}} \right)},
\end{equation}
where the $P_{\text{archival}}$ is adopted from \citet{feast_2008}.
As we merge data, the initial phase ($\phi^{0}$) may be modified by adjustments to the period, {\tt HJD$_{\text{max}}$}, 
 and inclusion of higher-order terms, which are fixed for an individual star.
The final phase ($\phi$) for any given data point is defined as:
\begin{equation} \label{eq:finalphase}
\phi_{i} = \frac{\vartheta_{i}}{P_{\text{final}}} - \rm{int}{\left( \frac{\vartheta_{i}}{P_{\text{final}}}\right)} + \zeta \left( \frac{\vartheta_{i}}{365.25} \right)^2,
\end{equation}
where $P_{\text{final}}$ is the final period,
 and $\zeta$ is an optional term (quadratic in $\vartheta$) that is used to describe
 changes in period in recent times, where $\zeta=0$ for a star with a stable period. 
The {\tt HJD$_{\text{max}}$} measured from our TMMT data and $\zeta$ and $P_{\text{final}}$, determined by the analyses 
 to follow, are given in Table \ref{tab:sample} for each of the stars in our sample. 

Our goal for this work is to build multi-wavelength light curves, and as such 
 our goal for phasing is to make all of the data sets for a given star conform to a single set of
 {\tt HJD$_{\text{max}}$}, $P_{\text{final}}$ and $\zeta$ that self-consistently phase all of the data-sets for a star (see Appendix \ref{app:indstars}). 
While this seems straightforward, in practice it is quite difficult
 owing both (i) to the nature of the observational data and (ii) to the nature of finding phasing solutions. 

Based on the sampling, each observational data set can be placed into one of three categories:
\begin{itemize}
 \item {\sc Case 1} ---well sampled light curves,
 \item {\sc Case 2} ---sparse coverage (or single points) for which there is a contemporaneous data set in the previous category (sparse data become `locked' to the data in {\sc Case 1}), and 
 \item {\sc Case 3} ---sparse coverage (or single points) with wide time baselines from well sampled curves. 
\end{itemize}
{\sc Case 1} and {\sc Case 2} can be analyzed and evaluated in the O-C diagram, which compares the observed (O) and computed or predicted (C) time of maximum or minimum light
 as a function of time (here we will use $\vartheta$).
The {\sc Case 2} data-sets become `locked' to their contemporaneous {\sc Case 1} data sets.
Usually the well sampled data sets can be merged by visual examination of their light curves. 
For {\sc Case 3}, the ephemerides for the majority of the well sampled data must be complete
 (e.g., the analyses for {\sc Case 1} and {\sc Case 2}) before the data set can be fully evaluated for
  consistency with the phased data sets. 
Usually, the {\sc Case 3} data set is `locked' to the nearest GEOS maxima observation and 
 phased to other data via small shifts in $\phi$.  

\subsection{The O-C Diagram}\label{ssec:ocdiagram}
Full evaluation of the ephemerides occurs within the context of the O-C diagram \citep[see Figure 3 of][for good demonstrations of various behaviors]{ocdiagrams}. 
{ O-C diagrams have a long history, beginning with \citet{luyten1921} and \citet{eddington1929}, and are utilized for a number of time-domain topics in astronomy. 
An excellent general introduction to O-C diagrams and their application for various science goals, as well as detailed discussion of misuse of such diagrams, is provided by \citet{sterken2005} and we refer the interested reader to that text. 
We now describe our use of the O-C diagram in the context of the goals of this work.}

The O-C diagrams for stars in our study could be classified into four characteristic behaviors, which are demonstrated in the panels of Figure \ref{fig:phase}:
 flat, linear, quadratic, and chaotic/jittery. {These behaviors are applied to when the the data directly used in this study were taken. DX Del (Figure \ref{fig:phase}(b)), for example, was flat and then became (positive) linear; the linear portion applies to all available data analyzed in this work.} 
None of the stars exhibited high-order periodic behavior that could be 
 associated with a close companion or other complicated physical scenario 
 \citep[for examples of these cases see discussions in][]{sterken2005,ocdiagrams}.
We discuss the implications for each of the situations in the sections to follow. 

\medbreak
\noindent {\sc Flat O-C Diagram}. If the data sets are 
{in phase} with the 
 literature period and TMMT {\tt HJD$_{\text{max}}$}, then the O-C diagram
 will show flat behavior as in the example in Figure \ref{fig:phase}a. 
Physically, this means that the period itself has been stable over the time frame of the data set.
If the TMMT {\tt HJD$_{\text{max}}$} is correct, then $\langle O-C \rangle=0$;
if it is incorrect, then there will be a zero-point offset. 
To reconcile, we adjust {\tt HJD$_{\text{max}}$},
 where a positive (negative) offset implies that the observed {\tt HJD$_{\text{max}}$} is 
 occurs later (earlier) than the value predicted by Equation \ref{eq:initialphase}.
{\tt HJD$_{\text{max}}$} is then adjusted such that the $\langle O-C \rangle=0$. 

\medbreak
\noindent {\sc Linear O-C Diagram}. If the data sets show linear 
 behavior (with nonzero slope) in the O-C diagram then observed maxima occur earlier (later) than predicted by Equation \ref{eq:initialphase}.
This is typically an indication that the period is incorrect
 in a way that accumulates over time, i.e., a constant difference between the true period and that initially used for phasing. 
The magnitude of the slope provides the amount of period mismatch
 and the sign of the slope indicates whether the period
 should be lengthened (negative) or shortened (positive). 

\medbreak
\noindent {\sc Quadratic O-C Diagram}. RRL with historical or 
 current constant period changes (most likely due to evolution) will have parabolic
 behavior in the O-C diagram.
 An upward (downward) parabola represents 
  a period that is lengthening (shortening) at a constant rate over time.
 Since the period is still evolving, we describe the evolution
  of the period with an additional term in lieu of providing
  the period for the current epoch
  (if the period became constant, then we would see discontinuity 
   from a parabola to linear). 
 An example is given in Figure \ref{fig:phase}c.
 The quadratic shape can be fit, resulting
  in an additional coefficient for phasing the 
  data, which we call $\zeta$ in Equation \ref{eq:initialphase}.
 Values of $\zeta$ are given in Table \ref{tab:sample},
  with `no data' indicating that no quadratic term was required. 
 
\medbreak
\noindent {\sc Chaotic/Jittery O-C Diagram}. Chaotic/jittery O-C diagrams
 could have many causes, including unresolved high-order variations due to companions, sudden period changes due to stellar evolution or other physical processes, typographical or computational errors in literature observations, and/or a combination of effects that cannot be identified individually \citep[see some examples of individual effects in][]{ocdiagrams}. 
{ Additionally, both sudden and prolonged chaotic and/or nonlinear effects could have causes unrelated to the physical evolution of the star.}
Merging data in these cases is quite complex.
Our general approach is to fit only the recent behavior in the O-C diagram to adjust the ephemerides
 (the last decade is usually covered by GEOS).
Older data sets are then treated individually, 
 often requiring individual phase shifts for merging. 
An example chaotic/jittery O-C diagram is given in Figure \ref{fig:phase}d { and for this demonstration we use RR Lyr itself, which appears chaotic/jittery in this visualization because of its strong Blazhko effect with a variable period (see Section \ref{rr_lyr} for details).}

\subsection{Final Ephemerides}\label{ssec:phasingsum} 
Making the data sets align to a single set of ephemerides is a multi-step process.
Data are converted to an initial phase ($\phi_0$) 
 based on the literature period ($P_{\text{archival}}$)
 and TMMT {\tt HJD$_{\text{max}}$} using Equation \ref{eq:initialphase}. 
 Iteration on the parameters occurs via visual inspection of the light curves from multiple data sets and the O-C Diagram.
{\sc Case 1} and {\sc Case 2} data-sets provide the most leverage on the ephemerides and are, as such, merged first, with {\sc Case 3} data-sets being tied to the closest GEOS epoch and folded in last.
Flat, linear, and quadratic O-C behaviors constrain adjustments to the ephemerides, which are also evaluated in the light curves.
An additional term, $\zeta$, may be added as in Equation \ref{eq:finalphase} to describe period changes in a quadratic O-C diagram. 
Final ephemerides are reported in Table \ref{tab:sample}, with some star-specific notes for chaotic/jittery O-C diagrams included in Appendix \ref{app:indstars}. 

Our final data sets report the final derived phase for each data point ($\phi_{i}$) using our adopted ephemerides as well as the original {\sc HJD} of observation.
Our process has generally preserved phase differences between filters.
We note that our goal in this process was to build multi-wavelength light curves for the eventual multi-wavelength calibration of period-luminosity relationships and not to find the highest-fidelity ephemerides for these stars.
Thus, while our solutions are adequate for our goal (as will be shown in the next section), they may not be unique solutions and may require further adjustment for other applications.


\section{Light Curves and Mean Magnitudes} \label{sec:meanmags}

The light curves constructed from the newly acquired TMMT data and phased archival data are shown in Fig \ref{fig:examples}. 
The complete figure set (55 images) is available in Appendix \ref{app:indstars}.  
The process of creating a light curve through the data using the \gloess{} technique is described in the following sections.  
From the evenly sampled \gloess{} light curve the intensity mean magnitude is determined as a simple average of the intensity of the \gloess{} data points and converted to a magnitude.  
Table \ref{tab:data} contains each new TMMT measurement along with all phased archival data included in this study.  
The full table is available in the online journal and a portion is shown here for form and content.   

\begin{figure*}
\begin{center}
\includegraphics[width=0.45\textwidth]{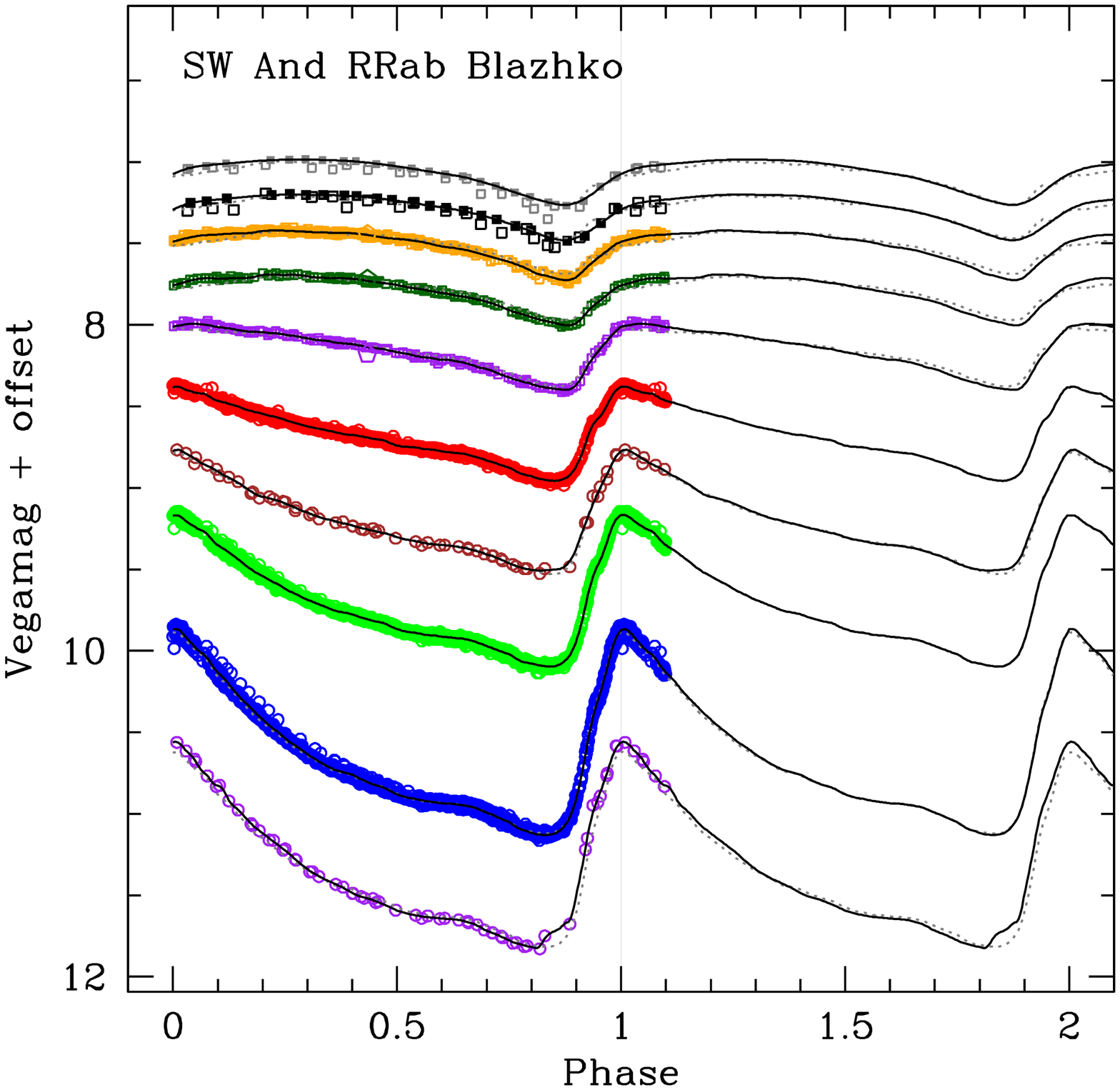} 
\includegraphics[width=0.45\textwidth]{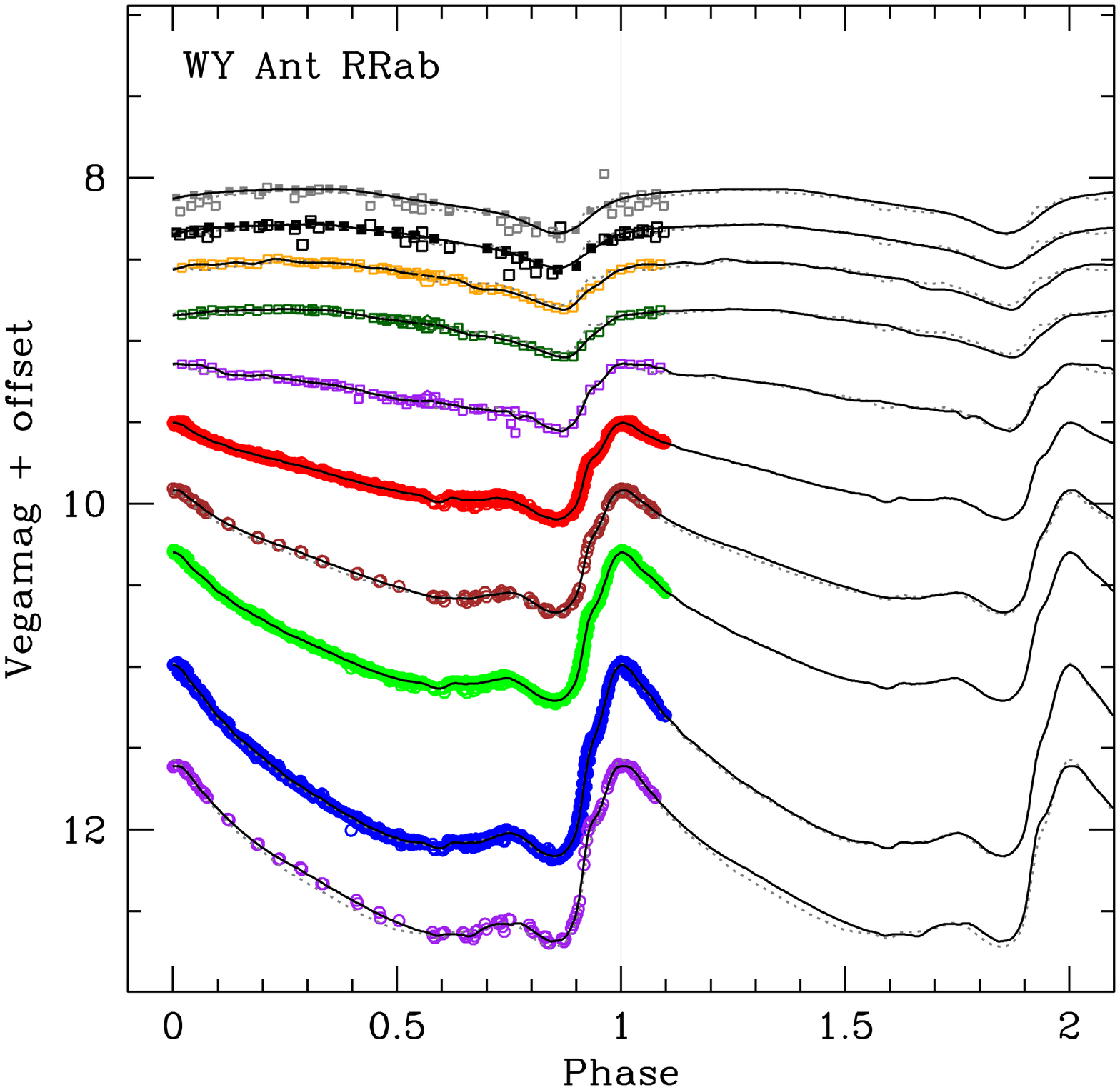} 
\includegraphics[width=0.45\textwidth]{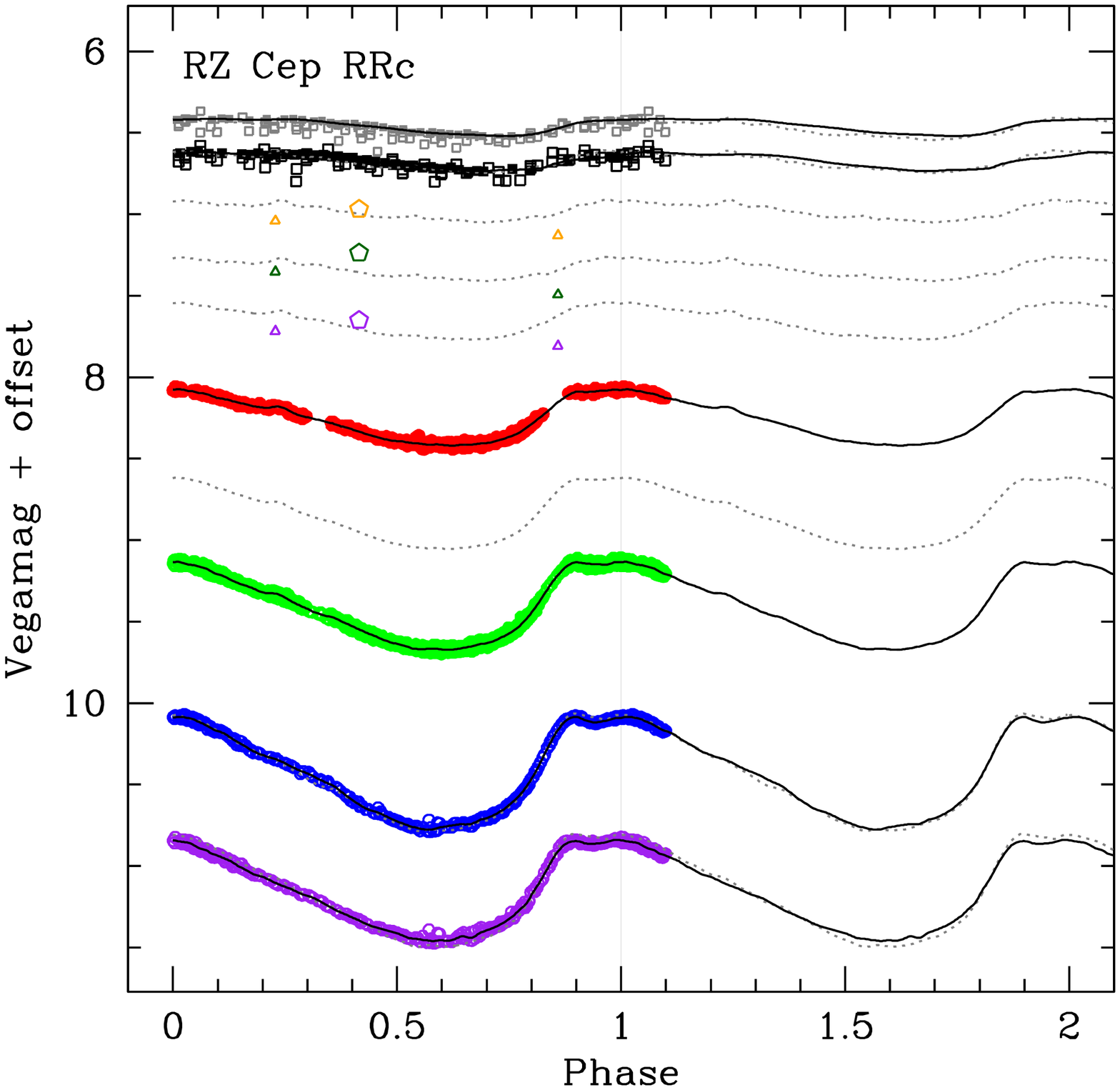}
\includegraphics[width=0.45\textwidth]{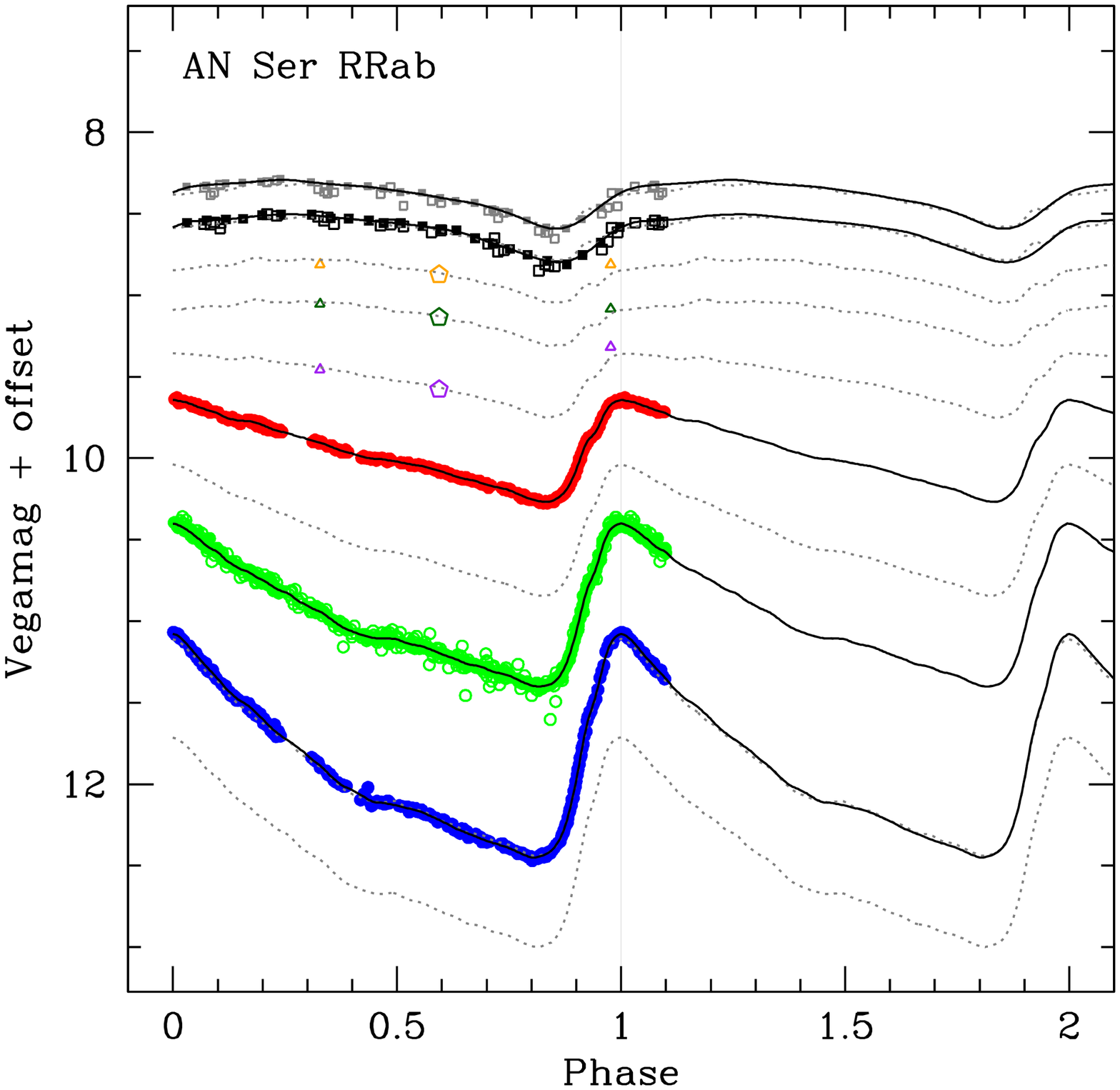}
\includegraphics[width=0.45\textwidth]{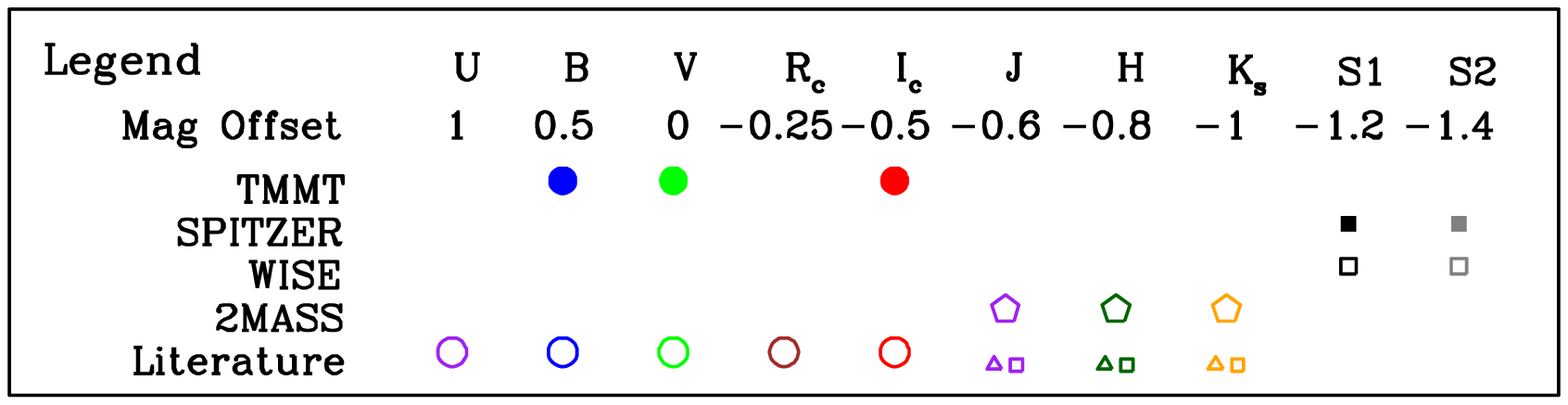}
\caption{ \label{fig:examples}
Example light curves from the final data sets. Data and \gloess{} fits for the RRab types SW And (top left) and WY Ant (top right) and the RRc types RZ Cep (bottom left) and T Sex (bottom right). 
 From top to bottom in each panel, 
  the bands are $S2$ (gray), $S1$ (black),
  $K_s$ (orange), $H$ (dark green), $J$ (purple),
  $I_C$ (red), $R_C$ (brown), $V$ (light green),
  $B$ (blue), and $U$ (purple). Filled points are newly acquired data from TMMT (circles) and \spitzer{} (squares).  All open symbols are adopted from the literature.
 Large open pentagons are $J$, $H$, $K_s$ from 2MASS, open triangles are NIR data from \cite{fernley_1993} and open squares are NIR data from the remaining literature sources.  
 For the MIR data, open symbols are from \emph{WISE} and filled symbols are from \spitzer{}.
 The \gloess{} light curve fit for each band is shown as the solid line {($3\sigma$ outliers were rejected during creation of the \gloess{} light curve). The dotted line is constructed from the $V$ and $I$ \gloess{} light curves for each band. The process of generating this faux curve and using it as a template is the topic of a future paper, but is briefly demonstrated here.} } 
\end{center}
\end{figure*}
\figsetstart
\figsetnum{4}
\figsettitle{Multi-Wavelength phased RR Lyrae light curves}
%
\figsetgrpstart
\figsetgrpnum{4.1}
\figsetgrptitle{SW And}
\figsetplot{0.eps}
\figsetgrpnote{Phased data and GLOESS light curves for SW And. \label{SW_And}}
\figsetgrpend

\figsetgrpstart
\figsetgrpnum{4.2}
\figsetgrptitle{XX And}
\figsetplot{2.eps}
\figsetgrpnote{Phased data and GLOESS light curves for XX And. \label{XX_And}}
\figsetgrpend

\figsetgrpstart
\figsetgrpnum{4.3}
\figsetgrptitle{WY Ant}
\figsetplot{4.eps}
\figsetgrpnote{Phased data and GLOESS light curves for WY Ant. \label{WY_Ant}}
\figsetgrpend

\figsetgrpstart
\figsetgrpnum{4.4}
\figsetgrptitle{X Ari}
\figsetplot{6.eps}
\figsetgrpnote{Phased data and GLOESS light curves for X Ari. \label{X_Ari}}
\figsetgrpend

\figsetgrpstart
\figsetgrpnum{4.5}
\figsetgrptitle{AE Boo}
\figsetplot{8.eps}
\figsetgrpnote{Phased data and GLOESS light curves for AE Boo. \label{AE_Boo}}
\figsetgrpend

\figsetgrpstart
\figsetgrpnum{4.6}
\figsetgrptitle{ST Boo}
\figsetplot{10.eps}
\figsetgrpnote{Phased data and GLOESS light curves for ST Boo. \label{ST_Boo}}
\figsetgrpend

\figsetgrpstart
\figsetgrpnum{4.7}
\figsetgrptitle{TV Boo}
\figsetplot{14.eps}
\figsetgrpnote{Phased data and GLOESS light curves for TV Boo. \label{TV_Boo}}
\figsetgrpend

\figsetgrpstart
\figsetgrpnum{4.8}
\figsetgrptitle{UY Boo}
\figsetplot{16.eps}
\figsetgrpnote{Phased data and GLOESS light curves for UY Boo. \label{UY_Boo}}
\figsetgrpend

\figsetgrpstart
\figsetgrpnum{4.9}
\figsetgrptitle{ST CVn}
\figsetplot{18.eps}
\figsetgrpnote{Phased data and GLOESS light curves for ST CVn. \label{ST_CVn}}
\figsetgrpend

\figsetgrpstart
\figsetgrpnum{4.10}
\figsetgrptitle{UY Cam}
\figsetplot{20.eps}
\figsetgrpnote{Phased data and GLOESS light curves for UY Cam. \label{UY_Cam}}
\figsetgrpend

\figsetgrpstart
\figsetgrpnum{4.11}
\figsetgrptitle{YZ Cap}
\figsetplot{24.eps}
\figsetgrpnote{Phased data and GLOESS light curves for YZ Cap. \label{YZ_Cap}}
\figsetgrpend

\figsetgrpstart
\figsetgrpnum{4.12}
\figsetgrptitle{RZ Cep}
\figsetplot{26.eps}
\figsetgrpnote{Phased data and GLOESS light curves for RZ Cep. \label{RZ_Cep}}
\figsetgrpend

\figsetgrpstart
\figsetgrpnum{4.13}
\figsetgrptitle{RR Cet}
\figsetplot{28.eps}
\figsetgrpnote{Phased data and GLOESS light curves for RR Cet. \label{RR_Cet}}
\figsetgrpend

\figsetgrpstart
\figsetgrpnum{4.14}
\figsetgrptitle{CU Com}
\figsetplot{30.eps}
\figsetgrpnote{Phased data and GLOESS light curves for CU Com. \label{CU_Com}}
\figsetgrpend

\figsetgrpstart
\figsetgrpnum{4.15}
\figsetgrptitle{RV CrB}
\figsetplot{33.eps}
\figsetgrpnote{Phased data and GLOESS light curves for RV CrB. \label{RV_CrB}}
\figsetgrpend

\figsetgrpstart
\figsetgrpnum{4.16}
\figsetgrptitle{W Crt}
\figsetplot{35.eps}
\figsetgrpnote{Phased data and GLOESS light curves for W Crt. \label{W_Crt}}
\figsetgrpend

\figsetgrpstart
\figsetgrpnum{4.17}
\figsetgrptitle{UY Cyg}
\figsetplot{38.eps}
\figsetgrpnote{Phased data and GLOESS light curves for UY Cyg. \label{UY_Cyg}}
\figsetgrpend

\figsetgrpstart
\figsetgrpnum{4.18}
\figsetgrptitle{XZ Cyg}
\figsetplot{40.eps}
\figsetgrpnote{Phased data and GLOESS light curves for XZ Cyg. \label{XZ_Cyg}}
\figsetgrpend

\figsetgrpstart
\figsetgrpnum{4.19}
\figsetgrptitle{DX Del}
\figsetplot{42.eps}
\figsetgrpnote{Phased data and GLOESS light curves for DX Del. \label{DX_Del}}
\figsetgrpend

\figsetgrpstart
\figsetgrpnum{4.20}
\figsetgrptitle{SU Dra}
\figsetplot{44.eps}
\figsetgrpnote{Phased data and GLOESS light curves for SU Dra. \label{SU_Dra}}
\figsetgrpend

\figsetgrpstart
\figsetgrpnum{4.21}
\figsetgrptitle{SW Dra}
\figsetplot{48.eps}
\figsetgrpnote{Phased data and GLOESS light curves for SW Dra. \label{SW_Dra}}
\figsetgrpend

\figsetgrpstart
\figsetgrpnum{4.22}
\figsetgrptitle{CS Eri}
\figsetplot{50.eps}
\figsetgrpnote{Phased data and GLOESS light curves for CS Eri. \label{CS_Eri}}
\figsetgrpend

\figsetgrpstart
\figsetgrpnum{4.23}
\figsetgrptitle{RX Eri}
\figsetplot{52.eps}
\figsetgrpnote{Phased data and GLOESS light curves for RX Eri. \label{RX_Eri}}
\figsetgrpend

\figsetgrpstart
\figsetgrpnum{4.24}
\figsetgrptitle{SV Eri}
\figsetplot{54.eps}
\figsetgrpnote{Phased data and GLOESS light curves for SV Eri. \label{SV_Eri}}
\figsetgrpend

\figsetgrpstart
\figsetgrpnum{4.25}
\figsetgrptitle{RR Gem}
\figsetplot{57.eps}
\figsetgrpnote{Phased data and GLOESS light curves for RR Gem. \label{RR_Gem}}
\figsetgrpend

\figsetgrpstart
\figsetgrpnum{4.26}
\figsetgrptitle{TW Her}
\figsetplot{60.eps}
\figsetgrpnote{Phased data and GLOESS light curves for TW Her. \label{TW_Her}}
\figsetgrpend

\figsetgrpstart
\figsetgrpnum{4.27}
\figsetgrptitle{VX Her}
\figsetplot{62.eps}
\figsetgrpnote{Phased data and GLOESS light curves for VX Her. \label{VX_Her}}
\figsetgrpend

\figsetgrpstart
\figsetgrpnum{4.28}
\figsetgrptitle{SV Hya}
\figsetplot{64.eps}
\figsetgrpnote{Phased data and GLOESS light curves for SV Hya. \label{SV_Hya}}
\figsetgrpend

\figsetgrpstart
\figsetgrpnum{4.29}
\figsetgrptitle{V Ind}
\figsetplot{66.eps}
\figsetgrpnote{Phased data and GLOESS light curves for V Ind. \label{V_Ind}}
\figsetgrpend

\figsetgrpstart
\figsetgrpnum{4.30}
\figsetgrptitle{BX Leo}
\figsetplot{68.eps}
\figsetgrpnote{Phased data and GLOESS light curves for BX Leo. \label{BX_Leo}}
\figsetgrpend

\figsetgrpstart
\figsetgrpnum{4.31}
\figsetgrptitle{RR Leo}
\figsetplot{70.eps}
\figsetgrpnote{Phased data and GLOESS light curves for RR Leo. \label{RR_Leo}}
\figsetgrpend

\figsetgrpstart
\figsetgrpnum{4.32}
\figsetgrptitle{TT Lyn}
\figsetplot{72.eps}
\figsetgrpnote{Phased data and GLOESS light curves for TT Lyn. \label{TT_Lyn}}
\figsetgrpend

\figsetgrpstart
\figsetgrpnum{4.33}
\figsetgrptitle{RR Lyr}
\figsetplot{74.eps}
\figsetgrpnote{Phased data and GLOESS light curves for RR Lyr. \label{RR_Lyr}}
\figsetgrpend

\figsetgrpstart
\figsetgrpnum{4.34}
\figsetgrptitle{RV Oct}
\figsetplot{79.eps}
\figsetgrpnote{Phased data and GLOESS light curves for RV Oct. \label{RV_Oct}}
\figsetgrpend

\figsetgrpstart
\figsetgrpnum{4.35}
\figsetgrptitle{UV Oct}
\figsetplot{83.eps}
\figsetgrpnote{Phased data and GLOESS light curves for UV Oct. \label{UV_Oct}}
\figsetgrpend

\figsetgrpstart
\figsetgrpnum{4.36}
\figsetgrptitle{AV Peg}
\figsetplot{85.eps}
\figsetgrpnote{Phased data and GLOESS light curves for AV Peg. \label{AV_Peg}}
\figsetgrpend

\figsetgrpstart
\figsetgrpnum{4.37}
\figsetgrptitle{BH Peg}
\figsetplot{87.eps}
\figsetgrpnote{Phased data and GLOESS light curves for BH Peg. \label{BH_Peg}}
\figsetgrpend

\figsetgrpstart
\figsetgrpnum{4.38}
\figsetgrptitle{DH Peg}
\figsetplot{90.eps}
\figsetgrpnote{Phased data and GLOESS light curves for DH Peg. \label{DH_Peg}}
\figsetgrpend

\figsetgrpstart
\figsetgrpnum{4.39}
\figsetgrptitle{RU Psc}
\figsetplot{92.eps}
\figsetgrpnote{Phased data and GLOESS light curves for RU Psc. \label{RU_Psc}}
\figsetgrpend

\figsetgrpstart
\figsetgrpnum{4.40}
\figsetgrptitle{BB Pup}
\figsetplot{94.eps}
\figsetgrpnote{Phased data and GLOESS light curves for BB Pup. \label{BB_Pup}}
\figsetgrpend

\figsetgrpstart
\figsetgrpnum{4.41}
\figsetgrptitle{HK Pup}
\figsetplot{96.eps}
\figsetgrpnote{Phased data and GLOESS light curves for HK Pup. \label{HK_Pup}}
\figsetgrpend

\figsetgrpstart
\figsetgrpnum{4.42}
\figsetgrptitle{RU Scl}
\figsetplot{98.eps}
\figsetgrpnote{Phased data and GLOESS light curves for RU Scl. \label{RU_Scl}}
\figsetgrpend

\figsetgrpstart
\figsetgrpnum{4.43}
\figsetgrptitle{SV Scl}
\figsetplot{100.eps}
\figsetgrpnote{Phased data and GLOESS light curves for SV Scl. \label{SV_Scl}}
\figsetgrpend

\figsetgrpstart
\figsetgrpnum{4.44}
\figsetgrptitle{AN Ser}
\figsetplot{102.eps}
\figsetgrpnote{Phased data and GLOESS light curves for AN Ser. \label{AN_Ser}}
\figsetgrpend

\figsetgrpstart
\figsetgrpnum{4.45}
\figsetgrptitle{AP Ser}
\figsetplot{105.eps}
\figsetgrpnote{Phased data and GLOESS light curves for AP Ser. \label{AP_Ser}}
\figsetgrpend

\figsetgrpstart
\figsetgrpnum{4.46}
\figsetgrptitle{T Sex}
\figsetplot{107.eps}
\figsetgrpnote{Phased data and GLOESS light curves for T Sex. \label{T_Sex}}
\figsetgrpend

\figsetgrpstart
\figsetgrpnum{4.47}
\figsetgrptitle{V0440 Sgr}
\figsetplot{109.eps}
\figsetgrpnote{Phased data and GLOESS light curves for V0440 Sgr. \label{V0440_Sgr}}
\figsetgrpend

\figsetgrpstart
\figsetgrpnum{4.48}
\figsetgrptitle{V0675 Sgr}
\figsetplot{111.eps}
\figsetgrpnote{Phased data and GLOESS light curves for V0675 Sgr. \label{V0675_Sgr}}
\figsetgrpend

\figsetgrpstart
\figsetgrpnum{4.49}
\figsetgrptitle{MT Tel}
\figsetplot{113.eps}
\figsetgrpnote{Phased data and GLOESS light curves for MT Tel. \label{MT_Tel}}
\figsetgrpend

\figsetgrpstart
\figsetgrpnum{4.50}
\figsetgrptitle{AM Tuc}
\figsetplot{115.eps}
\figsetgrpnote{Phased data and GLOESS light curves for AM Tuc. \label{AM_Tuc}}
\figsetgrpend

\figsetgrpstart
\figsetgrpnum{4.51}
\figsetgrptitle{AB UMa}
\figsetplot{117.eps}
\figsetgrpnote{Phased data and GLOESS light curves for AB UMa. \label{AB_UMa}}
\figsetgrpend

\figsetgrpstart
\figsetgrpnum{4.52}
\figsetgrptitle{RV UMa}
\figsetplot{122.eps}
\figsetgrpnote{Phased data and GLOESS light curves for RV UMa. \label{RV_UMa}}
\figsetgrpend

\figsetgrpstart
\figsetgrpnum{4.53}
\figsetgrptitle{SX UMa}
\figsetplot{124.eps}
\figsetgrpnote{Phased data and GLOESS light curves for SX UMa. \label{SX_UMa}}
\figsetgrpend

\figsetgrpstart
\figsetgrpnum{4.54}
\figsetgrptitle{TU UMa}
\figsetplot{127.eps}
\figsetgrpnote{Phased data and GLOESS light curves for TU UMa. \label{TU_UMa}}
\figsetgrpend

\figsetgrpstart
\figsetgrpnum{4.55}
\figsetgrptitle{UU Vir}
\figsetplot{129.eps}
\figsetgrpnote{Phased data and GLOESS light curves for UU Vir. \label{UU_Vir}}
\figsetgrpend

\figsetend
\begin{deluxetable*}{lccccccc}
\tablewidth{0pt}
\tabletypesize{\scriptsize}
\tablecaption{TMMT Photometry and Phased Archival Data
\label{tab:data}}
\tablehead{
\colhead{Star}           & \colhead{Filter}      &
\colhead{mag}          & \colhead{$\sigma_{phot}$}  &
\colhead{$\sigma_{sys}$}          & \colhead{HMJD}  & \colhead{{\bf Phase ($\phi$)}}  &
\colhead{Reference} 
}
\startdata
    SW And  &   I &  9.170 &  0.010 &  0.009 &   56549.3206 & 0.390  & 0 \\
    SW And  &   I &  9.168 &  0.010 &  0.009 &   56549.3216 & 0.392  & 0 \\
    SW And  &   I &  9.173 &  0.010 &  0.009 &   56549.3222 & 0.394  & 0 \\
    SW And  &   I &  9.174 &  0.010 &  0.009 &   56549.3229 & 0.395  & 0 \\
    SW And  &   I &  9.171 &  0.010 &  0.009 &   56549.3235 & 0.397  & 0 \\   
\enddata
\tablecomments{ The Heliocentric Modified Julian Day (HMJD = HJD - 2400000.5) is provided.  The photometric error for each measurement is included as well as the systematic error in the zero-point determination.
Table \ref{tab:data} is published in its entirety in the machine-readable format.
      A portion is shown here for guidance regarding its form and content.}
\tablerefs{
(0)~TMMT This work; (1)~\spitzer{} This work; (4)~\citet{skillen_1993jhk}; (5)~\citet{Barnes_1992}; (7)~\citet{liu_1989}; (8)~\citet{liu_1989}; (9)~\citet{Barcza_2014}; (10)~\citet{Pac_1965_2}; (11)~2MASS \citet{skrutskie_2006};
(17)~ASAS \citet{pojmanski_1997};
(15)~\citet{jones_1992};
(19)~IBVS \citet{Broglia1992};
(31)~\citet{fernley_1990};
(41)~\citet{fernley_1989};
(98)~\citet{Clementini_1990};
(99)~\citet{clementini_2000};
(999)~TMMT modified for Blazhko effect. 
}
\end{deluxetable*}
\subsection{GLOESS Light Curve Fitting} \label{sec:gloess}

Nonparametric kernel regression and local polynomial fitting have a long history, dating as far back as \citet{macaulay1931}; in particular they have been extensively applied to the analysis of time series data.
More recently popularized and developed by \citet{cleveland1979} and \citet{cleveland1988}, this method has been given the acronym {LOESS} (standing for LOcal regrESSion), or alternatively and less frequently {\sc LOWESS} (standing for LOcally WEighted Scatterplot Smoothing). 
In either event, a finite-sized, moving window (a kernel of finite support) was used to select data, which were then used in a polynomial regression to give a single interpolation point at the center of the adopted kernel (usually uniform or triangular windows). 
The kernel was then moved by some interpolation interval determined by the user and the process repeated until the entire data set was scanned. 
Instabilities would occur when the window was smaller than the largest gaps between consecutive data points. 
To eliminate this possibility one of us (BFM) introduced \gloess, a Gaussian-windowed LOcal regrESSion  method, first used by \citet{persson_2004} to fit Cepheid light curves. 
In its simplest form \gloess{} penalized the data, 
 both ahead of and behind the center point of the window, by their Gaussian-weighted distance 
 quadratically convolved with their individual statistical errors.
Instabilities are guaranteed to be avoided given that all data contribute to the polynomial regressions at every step.  

\gloess{} light curves were created for our stars for each band independently, the goal being to create uniformly interpolated light curves from nonuniformly sampled data.  
In this way, for example, color curves and/or other more complicated combinations of colors and magnitudes can be derived from multi-wavelength data sets that were collected in different bands at disparate and non-overlapping epochs \citep[see for example the application in][]{freedman2010b}. 
In our particular implementation of the \gloess{} formalism the width of the window, $\Delta x$ (usually phase), 
 could be either set to a constant or allowed to vary as the density of available data points changes.
The latter allows for finer detail to be captured in the interpolation in regions where there are more densely populated data points. 
Our implementation has been extended further by also weighting in $\Delta y$ (magnitude).  
By doing a first-pass linear interpolation the data points are weighted by an additional factor of $e^{-\Delta_y^2/\sigma_y^2}$.  
In this way, regions near the top and bottom of a steeply rising feature are ``shielded'' from each other. 
In this case, $\sigma_y$ is the scatter in the data within the $\Delta x$ window.  {Outliers are clipped by performing a second iteration and rejecting points outside $3\sigma$.}
For some stars, it proved useful to partition the data into discrete regions to further isolate steeply changing features from one another.

The results of the \gloess{} fitting are shown for each of the light curves given in 
 Figure \ref{fig:examples} as the solid gray lines. 
In comparison to the data points (shown for $0 < \phi < 1.0$), 
 the \gloess{} curve (shown in isolation for $1.0 < \phi < 2.0$) 
 accurately traces the natural structure of the light curves
 with no a priori assumptions of that shape.
There are two {types of} cases in our current sample that require special attention.
These are (i) those stars exhibiting the Blazhko effect in our sampling
 (typically only visible in TMMT data) and (ii) those stars with large phase gaps. 

\subsubsection{Treatment of the Blazhko Effect}\label{sec:blaz}

The Blazhko effect is a modulation of the amplitude and shape of the{ RRL light curve}
 with periods ranging from a few to hundreds of days.
The search for a physical basis for the Blazhko effect
 remains unclear globally and is beyond the scope of this paper.  
 The average luminosity of Blazhko stars remains constant from cycle to cycle; however; data that nonuniformly sample the light curve from different Blazhko cycles may lead to an incorrect determination of the mean, biased in the direction where most of the data were obtained.  
For the purposes of this paper, we did the following:  If the observations could be separated cleanly into distinct Blazhko cycles, then the two cycles were shifted and scaled to one another if necessary for the purposes of \gloess{} fitting (for example, SV Hya and RV UMa). Data displayed in these figures are the original photometry {with the modified photometry displayed in gray}.  If the observations could {not readily} be separated, then the \gloess{} algorithm was left to average between the cycles. 

\subsubsection{Light-curve Partitioning}\label{sec:partitioning}

One feature of \gloess{} is that points ahead of or behind the current interpolation point are downweighted.  
In the optical bands, the light curves can have very rapid rise times, and even with the the weighting function sometimes the data on the ascending or descending side of the light curve still influence the local estimation of the others. 
Here we briefly discuss the general modification of \gloess{} for these situations. 
 
For the purposes of this work, the phase between minimum light and just past maximum light was often `partitioned' such that the points on either side of the maximum could not influence each other. 
This partitioning was particularly necessary for the optical light curves of RRab variables.
This was done by setting the weights of those points outside the fitting partition to zero.  {To prevent discontinuities at or near the partition point, data from within 0.02 in phase were allowed to contribute from the opposite side.  In this way, there were always data to interpolate and \gloess{} did not have to extrapolate.   }
 
Another feature of \gloess{} is that at each point of interpolation a quadratic function locally fits the weighted data.  
This low-order function has the advantage of not overfitting the data and generally changes slowly at adjacent interpolation points, thus providing a relatively smooth and continuous curve through the data in the end.  
At the timing of the `hump' in the ascending branch of the light curve for RRabs in the optical \citep[see][Figure 6 for a visualization]{chadid_2014}, the function is allowed to use the best $\chi^{2}$ result for a first-, second-, or third-order order polynomial.  
Because the `hump' happens so quickly there are often not enough data to capture the subtlety of the feature.  
Allowing for a third-order polynomial in this region avoids losing this and similar features in the light curves.  
RRcs are relatively smooth (in comparison to RRabs) and we always use an underlying quadratic function. 

\subsection{Light-curve Properties} \label{sec:lcproperties}

Mean magnitudes were determined by computing the mean intensity of the evenly sampled \gloess{} fit points to the light curve, then converting back to a magnitude. 
The light curve must be sampled with enough data points to capture all the nuances of the shape for an accurate mean; in our case we sampled with 256 data points spaced every 1/256 in phase. 
Generally, \gloess{} fits were determined only for those stars and bands that had more than 
 20 individual data points over a reasonable portion of phase (i.e., 20 data points
 only spanning $\phi\sim$~0.1 would not have a \gloess{} fit nor a mean magnitude).
 Table \ref{tab:gloess} contains each \gloess{}-generated light curve.  
The full table is available in the online journal and a portion is shown here for form and content. 

The random uncertainty of the \gloess{}-derived mean magnitude is simply the error
 on the mean of data points going into \gloess{} fitting.
Thus, stars with more data points will have a smaller uncertainity in \gloess{} mean magnitude. 
The systematic uncertainty is determined by the photometric transformations, either 
 in transforming our TMMT photometry onto an absolute system (see Figure \ref{fig:stdmag})
 or as reported in the literature and in transforming from other filter systems (as described in Section \ref{sec:data}). { The final reported error is $\sigma^2 = 1/\sum_{\rm{}}(1/\sigma_{phot}^{2}) + 1/\sum_{\rm{}}(1/\sigma_{sys}^2) $, where the sum over $\sigma_{sys}$ includes only the unique entries from each reference, i.e. it is not counted for every measurement.  }
These results are given in Table \ref{tab:meanmag}. 
A technique to better utilize sparsely sampled data will be presented in a future companion paper (R. L. Beaton et al. 2017, ~in preparation) and no mean magnitudes are reported here for data with too few measurements to construct a \gloess{} light curve.

In addition to the mean magnitudes, we provide 
 amplitudes ($a_{\lambda}$), rise times ($RT_{\lambda}$), and magnitudes at {\tt HJD$_{\text{max}}$} in Table  \ref{tab:lcpars} as measured from the \gloess{} light curves.
We define ($a_{\lambda}$) and ($RT_{\lambda}$) as the difference in magnitude and phase, respectively, between the minimum and maximum of the \gloess{} light curve.
We note that at the longer wavelengths these terms become less well defined due to the less prominent `saw-tooth' shape and overall smaller amplitudes, { both of which are typical changes for RRL stars at these passbands.}

\begin{deluxetable}{lccc}
\tablewidth{0pt}
\tabletypesize{\scriptsize}
\tablecaption{\gloess{} light curves.\label{tab:gloess}}
\tablehead{
\colhead{Star Name}       & \colhead{Filter}      &
\colhead{$\phi$}          & \colhead{\gloess{} Mag.}  
}
\startdata
    SW And  &   U & 0.00000 &  9.562 \\
    SW And  &   U & 0.00391 &  9.559 \\
    SW And  &   U & 0.00781 &  9.560 \\
    SW And  &   U & 0.01172 &  9.563 \\
    SW And  &   U & 0.01562 &  9.571 \\
\enddata
\tablecomments{Table \ref{tab:gloess} is published in its entirety in the machine-readable format.
      A portion is shown here for guidance regarding its form and content.}
\end{deluxetable}
\subsection{Comparison with Literature Values} \label{sec:comp}

\begin{figure}
\begin{centering}
\includegraphics[width=\columnwidth]{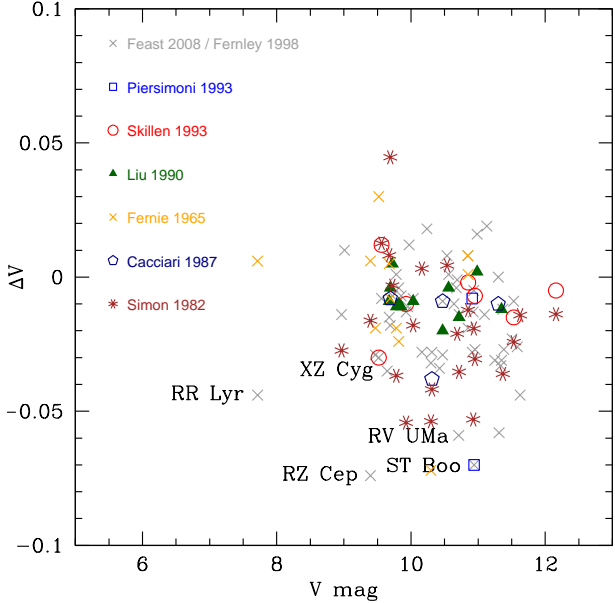}
\caption{Difference between the current intensity mean values <$V$> and values published in the literature as described in the text.}
\label{fig:comp}
\end{centering}
\end{figure}

The difference between the intensity means in the $V$ band determined here and literature values is shown in Figure \ref{fig:comp}.  In most cases $B$ photometry is available for comparison with respect to $V$ (i.e. $B-V$), but $I_C$ is not, which was part of the motivation for this work.   
Notable outliers are labeled in Figure \ref{fig:comp} and it is worth noting that three
 out of the five RRL with \hst{} parallaxes are considered outliers.  
Both RV UMa and ST Boo exhibit the Blazhko effect (our treatment of these stars was discussed in Section \ref{sec:blaz}) and differences are likely due to having sampled different parts of the Blazhko cycle.  

We compare our \gloess{}-derived apparent mean magnitudes for 53 stars to those presented in \citet{feast_2008}, which were originally derived in \cite{fernley_1998} from \hip{} Fourier fitting to $H_P$ magnitudes (converted to intensity).  
A correction was used in \cite{fernley_1998} to transform the RRL from the $H_P$ system onto the standard Johnson $V$ system by adopting an average color for each subtype of star; the corrections were $-0.09$~mag for the RRabs and $-0.06$~mag for the RRcs.  
Some of the scatter from \cite{feast_2008} and \cite{fernley_1998} in Figure \ref{fig:comp} is due to reddening, since the corrections were not based on apparent color but rather by adopting a mean color for each type of star.    
RZ Cep (type RRc), for example, 
 is highly reddened and has an apparent color redder than most RRabs (which should be intrinsically redder than RRcs).
 Thus RZ Cep's (and the others') color correction was likely underestimated when transforming from $H_P$ to $V$.  Both XZ Cyg and RR Lyr are also offset, perhaps due to their Blazhko cycle.  The mean magnitude from \cite{1965F} for RR Lyr is in agreement with the value determined here.  
For 51 stars in common (two rejected) we find an average offset of $\Delta V=-0.018\pm0.021$~mag.

\cite{Piersimoni1993} provide photometry for two stars (AB UMa and ST Boo).
There is good agreement on AB UMa, but the reported mean magnitude reported for ST Boo (a Blazhko star) differs by 0.07~mag. 

There are a number of stars with BW analysis; the results of  \citet{skillen_1993,skillen_1989} and \citet{fernley_1989,fernley_1990} are grouped together in Fig \ref{fig:comp} as "Skillen 1993." 
For the seven stars in common (one rejected), we find an average offset of $\Delta V = -0.005 \pm 0.009$ magnitudes.
W Crt was discussed in \cite{skillen_1993} as perhaps having an offset from observations between different telescopes.

We have 10 stars in common with \citet{liu_1990} and we we find an average offset of $\Delta V = -0.008 \pm 0.007$~mag.
For the 10 stars in common with \citet{1965F} we find an average offset of $\Delta V = -0.001 \pm 0.016$~mag.
For three stars in common with \citet[][we exclude V0440 Sgr]{cacciari_1987}, we find an average offset of $\Delta V = -0.009\pm0.001$~mag.

\cite{1982Simon} adopted photometry for a number of stars from earlier sources to generate Fourier fits to the light curve for each star.  
The fits were performed in magnitudes so the Fourier parameters provided were used to generate light curves from which the intensity-averaged magnitudes were determined. 
For the 23 stars in common with \citet{1982Simon}, we find an average offset of $\Delta V = -0.019\pm0.024$~mag.

{Overall, the average offsets in the Johnson $V$ band between the literature intensity mean values in the literature and our current results fall between 1\% and 2\%.  On average this would indicate that the current photometry is about 1\% brighter than previous estimates.  To check potential systematics, random standard stars observed with TMMT were processed like the RRL sample and the final mean magnitudes did not deviate from their standard values. Additionally, to check the \gloess{} method, literature data was passed through the \gloess{} algorithm and their final mean magnitudes agreed with their published values. The source of any systematic remains unclear and a full investigation into the potential systematics
involved in the complete assimilation of other data from the literature is 
beyond the scope of this paper. }


\section{Summary and Future Work} \label{sec:sum}

With the upcoming \gaia{} results, we will soon be able to calibrate the period--luminosity relationship directly using trigonometric parallaxes for a large sample of nearby Galactic RR Lyrae variables. 
In anticipation of the \gaia{} data releases, 
 we have prepared a data set spanning a wavelength range from 0.4 to 4.5 microns 
 in 10 individual photometric bands for a sample of 55 bright, nearby RR Lyrae variables
 that will be in the highest-precision \gaia{} sample.
Moreover, 53 of the 55 stars appear in the \hip{} catalog, and { a large fraction were a part of} the \gaia{} first data release\ with the \emph{Tycho-Gaia Astrometric Solution} \citep[TGAS;][]{michalik_2015,lindegren_2016}.
Our sample spans a representative range of RRL properties, containing both RRab and RRc type stars, 
 a wide range of metallicity, and several stars showing short-term and long-term Blazhko modulations.

In this paper, we described the TMMT, an automated, small-aperture facility designed to obtain high-precision, multi-epoch photometry for calibration sources.
We presented a multi-site and multi-year campaign 
 with the TMMT that produced well-sampled optical light curves
 for our (55-star) sample.
Additionally, we utilize MIR light curves obtained in the CRRP. 
Furthermore, we present an extensive literature search of the photometry
 to expand our phase coverage at all wavelengths.
We described our efforts to merge these data sets to conform to a single set of ephemerides 
 ({\tt HJD}$_{max}$, period, and higher-order terms)
 and explicitly include both our filter transformations and phasing solutions. 

With multi-wavelength merged data sets, we apply the \gloess{} technique to produce well-sampled, smoothed light curves for as many stars and bands as possible.
\gloess{} produces light curves that are not scaled templates or analytic functions,
 but are generated from a stars' actual data and thereby preserve the details of their often unique light-curve sub-structure. 
We describe adaptations of this technique required for application to stars
 with large amplitude modulations due to the Blazhko effect, and with other light -curve features that can present challenges.
The \gloess{} light curves are then used to determine high-precision intensity mean magnitudes and mean light-curve properties including amplitudes, rise times, and magnitudes at minimum and maximum light.

While our study is as complete as possible, many stars have observational data that 
 are currently not well sampled enough for the direct application of \gloess{}.
{ We have \gloess{} light curves for 22 (40\%) stars in the $U$, 
55 (100\%) stars in $B$, 55 (100\%) stars in $V$, 20 (36\%) stars in $R$, 55 (100\%) stars in $I$, 19 (35\%) stars in $J$, 9 (16\%) stars in $H$, 20 (36\%) stars in $K_s$, 55 (100\%) in [3.6], and 55 (100\%) in [4.5].
Our sample is particularly limited for the near-infrared}, 
 where high-cadence observations of bright stars are challenging or observationally expensive.
A companion paper will present a technique, schematically described in \citet{beaton_2016},
 that uses the TMMT-derived \gloess{} optical light curves presented in this work 
 to produce star-by-star predictive templates capable of making use of 
 single-phase or sparsely sampled data sets for RR Lyrae. 

\section*{Acknowledgments}
We thank the anonymous referee for helpful comments on the manuscript. 
We acknowledge helpful conversations with George Preston. 

This publication makes use of data products from the Two Micron All Sky Survey, which is a joint project of the University of Massachusetts and the Infrared Processing and Analysis Center/California Institute of Technology, funded by the National Aeronautics and Space Administration and the National Science Foundation.

This work is based (in part) on observations made with the \spitzer~ Space Telescope, which is operated by the Jet Propulsion Laboratory, California Institute of Technology under a contract with NASA.

This publication makes use of data products from the \emph{Wide-field Infrared Survey Explorer}, which is a joint project of the University of California, Los Angeles, and the Jet Propulsion Laboratory/California Institute of Technology, funded by the National Aeronautics and Space Administration.
This publication also makes use of data products from \emph{NEOWISE}, which is a project of the Jet Propulsion Laboratory/California Institute of Technology, funded by the Planetary Science Division of the National Aeronautics and Space Administration.

\facility{Spitzer (IRAC)}
\software{DAOPHOT \citep{stetson_1987}, DAOGROW \citep{stetson_1990}, PHOTCAL \citep{1993ASPC...52..479D}}

%
\begin{deluxetable*}{l r r r r r r r r r r  }
\tabletypesize{\tiny}
\tablecaption{Intensity Mean Magnitudes from \gloess{} light curves \label{tab:meanmag}}
\tablewidth{0pt}
\tablehead{
Name & \multicolumn{1}{c}{$U$} &  \multicolumn{1}{c}{$B$} & \multicolumn{1}{c}{$V$} & \multicolumn{1}{c}{$R_C$} & \multicolumn{1}{c}{$I_C$} & \multicolumn{1}{c}{$J$} & \multicolumn{1}{c}{$H$} & \multicolumn{1}{c}{$K_s$} & \multicolumn{1}{c}{[3.6]} & \multicolumn{1}{c}{[4.5]}
}
\startdata
SW And       & 10.287 0.020 & 10.097 0.006 & 9.692 0.006 & 9.433 0.020 & 9.169 0.008 & 8.757 0.020 & 8.590 0.013 & 8.511 0.009 & 8.485 0.009 & 8.472 0.008 \\
XX And       & \nodata \nodata & 11.018 0.009 & 10.676 0.009 & \nodata \nodata & 10.145 0.009 & \nodata \nodata & \nodata \nodata & \nodata \nodata & 9.409 0.009 & 9.384 0.008 \\
WY Ant       & 11.262 0.020 & 11.217 0.004 & 10.851 0.004 & 10.601 0.020 & 10.324 0.004 & 9.915 0.020 & 9.696 0.013 & 9.599 0.009 & 9.567 0.009 & 9.548 0.008 \\
X Ari       & 10.250 0.010 & 10.061 0.006 & 9.562 0.006 & 9.231 0.020 & 8.868 0.007 & 8.306 0.020 & 8.057 0.013 & 7.926 0.009 & 7.885 0.009 & 7.859 0.009 \\
AE Boo       & \nodata \nodata & 10.887 0.009 & 10.640 0.009 & \nodata \nodata & 10.254 0.009 & \nodata \nodata & \nodata \nodata & \nodata \nodata & 9.750 0.011 & 9.749 0.011 \\
ST Boo       & \nodata \nodata & 11.265 0.012 & 10.940 0.012 & \nodata \nodata & 10.484 0.012 & \nodata \nodata & \nodata \nodata & \nodata \nodata & 9.834 0.009 & 9.816 0.009 \\
TV Boo       & 11.230 0.020 & 11.179 0.009 & 10.986 0.009 & 10.838 0.020 & 10.652 0.009 & 10.294 0.009 & \nodata \nodata & 10.181 0.009 & 10.197 0.009 & 10.179 0.009 \\
UY Boo       & \nodata \nodata & 11.280 0.009 & 10.927 0.009 & \nodata \nodata & 10.428 0.009 & \nodata \nodata & \nodata \nodata & \nodata \nodata & 9.721 0.008 & 9.696 0.009 \\
ST CVn       & \nodata \nodata & 11.591 0.010 & 11.337 0.010 & \nodata \nodata & 10.949 0.010 & \nodata \nodata & \nodata \nodata & \nodata \nodata & 10.437 0.009 & 10.413 0.009 \\
UY Cam       & \nodata \nodata & 11.685 0.009 & 11.507 0.009 & \nodata \nodata & 11.207 0.028 & \nodata \nodata & \nodata \nodata & \nodata \nodata & 10.778 0.009 & 10.763 0.009 \\
YZ Cap       & 11.777 0.020 & 11.560 0.007 & 11.300 0.007 & 11.134 0.020 & 10.902 0.007 & \nodata \nodata & \nodata \nodata & \nodata \nodata & 10.337 0.009 & 10.321 0.009 \\
RZ Cep       & 10.143 0.020 & 9.907 0.020 & 9.396 0.011 & \nodata \nodata & 8.747 0.014 & \nodata \nodata & \nodata \nodata & \nodata \nodata & 7.871 0.008 & 7.858 0.009 \\
RR Cet       & 10.146 0.020 & 10.075 0.006 & 9.723 0.006 & 9.476 0.020 & 9.219 0.006 & 8.790 0.020 & \nodata \nodata & 8.518 0.009 & 8.501 0.009 & 8.489 0.008 \\
CU Com       & \nodata \nodata & 13.654 0.009 & 13.346 0.009 & \nodata \nodata & 12.873 0.009 & \nodata \nodata & \nodata \nodata & \nodata \nodata & 12.265 0.009 & 12.250 0.009 \\
RV CrB       & \nodata \nodata & 11.618 0.008 & 11.383 0.007 & \nodata \nodata & 11.012 0.007 & \nodata \nodata & \nodata \nodata & \nodata \nodata & 10.493 0.008 & 10.470 0.009 \\
W Crt       & 12.013 0.020 & 11.844 0.007 & 11.531 0.007 & 11.347 0.020 & 11.099 0.007 & 10.798 0.020 & 10.629 0.013 & 10.523 0.009 & 10.506 0.011 & 10.500 0.011 \\
UY Cyg       & \nodata \nodata & 11.515 0.017 & 11.096 0.017 & \nodata \nodata & 10.496 0.017 & \nodata \nodata & \nodata \nodata & \nodata \nodata & 9.709 0.009 & 9.688 0.008 \\
XZ Cyg       & \nodata \nodata & 9.903 0.008 & 9.645 0.008 & \nodata \nodata & 9.237 0.008 & \nodata \nodata & \nodata \nodata & \nodata \nodata & 8.657 0.009 & 8.639 0.008 \\
DX Del       & \nodata \nodata & 10.359 0.007 & 9.927 0.007 & \nodata \nodata & 9.367 0.007 & 8.997 0.020 & 8.818 0.013 & 8.709 0.009 & 8.650 0.009 & 8.637 0.008 \\
SU Dra       & 10.175 0.020 & 10.124 0.007 & 9.781 0.007 & 9.556 0.020 & 9.286 0.007 & 8.898 0.020 & \nodata \nodata & 8.635 0.009 & 8.598 0.008 & 8.580 0.009 \\
SW Dra       & 10.836 0.020 & 10.815 0.007 & 10.471 0.007 & 10.238 0.020 & 9.977 0.007 & \nodata \nodata & \nodata \nodata & \nodata \nodata & 9.303 0.009 & 9.285 0.009 \\
CS Eri       & \nodata \nodata & 9.244 0.006 & 9.010 0.006 & \nodata \nodata & 8.657 0.006 & \nodata \nodata & \nodata \nodata & \nodata \nodata & 8.125 0.008 & 8.109 0.009 \\
RX Eri       & 10.184 0.020 & 10.083 0.003 & 9.675 0.003 & 9.411 0.020 & 9.120 0.003 & \nodata \nodata & \nodata \nodata & 8.358 0.009 & 8.336 0.009 & 8.312 0.008 \\
SV Eri       & \nodata \nodata & 10.357 0.004 & 9.949 0.004 & \nodata \nodata & 9.379 0.004 & \nodata \nodata & \nodata \nodata & \nodata \nodata & 8.566 0.008 & 8.549 0.008 \\
RR Gem       & 11.912 0.020 & 11.689 0.011 & 11.349 0.011 & 11.122 0.020 & 10.874 0.011 & 10.485 0.020 & \nodata \nodata & 10.215 0.009 & 10.239 0.009 & 10.217 0.009 \\
TW Her       & \nodata \nodata & 11.554 0.005 & 11.249 0.005 & \nodata \nodata & 10.817 0.005 & \nodata \nodata & \nodata \nodata & \nodata \nodata & 10.234 0.009 & 10.210 0.009 \\
VX Her       & \nodata \nodata & 10.978 0.009 & 10.689 0.009 & \nodata \nodata & 10.244 0.009 & \nodata \nodata & \nodata \nodata & \nodata \nodata & 9.593 0.008 & 9.573 0.009 \\
SV Hya       & 10.941 0.021 & 10.849 0.006 & 10.538 0.006 & \nodata \nodata & 10.059 0.006 & \nodata \nodata & \nodata \nodata & \nodata \nodata & 9.367 0.008 & 9.348 0.009 \\
V Ind       & \nodata \nodata & 10.282 0.006 & 9.972 0.006 & 9.767 0.020 & 9.509 0.006 & \nodata \nodata & \nodata \nodata & \nodata \nodata & 8.849 0.009 & 8.830 0.009 \\
BX Leo       & \nodata \nodata & 11.831 0.006 & 11.584 0.006 & \nodata \nodata & 11.210 0.006 & \nodata \nodata & \nodata \nodata & \nodata \nodata & 10.678 0.009 & 10.670 0.009 \\
RR Leo       & 11.119 0.020 & 11.000 0.007 & 10.716 0.007 & 10.524 0.020 & 10.280 0.007 & 9.918 0.020 & \nodata \nodata & 9.659 0.009 & 9.658 0.008 & 9.630 0.008 \\
TT Lyn       & 10.297 0.020 & 10.217 0.011 & 9.853 0.011 & 9.605 0.020 & 9.318 0.011 & 8.896 0.020 & \nodata \nodata & 8.614 0.009 & 8.586 0.009 & 8.571 0.009 \\
RR Lyr       & 8.155 0.010 & 8.101 0.006 & 7.716 0.006 & \nodata \nodata & 7.206 0.007 & 6.746 0.009 & 6.598 0.013 & 6.496 0.009 & 6.470 0.009 & 6.461 0.009 \\
RV Oct       & \nodata \nodata & 11.386 0.007 & 10.953 0.007 & 10.680 0.020 & 10.335 0.007 & 9.863 0.020 & 9.610 0.013 & 9.490 0.009 & 9.492 0.011 & 9.480 0.011 \\
UV Oct       & \nodata \nodata & 9.844 0.007 & 9.471 0.007 & \nodata \nodata & 8.940 0.007 & \nodata \nodata & \nodata \nodata & \nodata \nodata & 8.180 0.009 & 8.167 0.009 \\
AV Peg       & 11.091 0.020 & 10.862 0.008 & 10.471 0.008 & 10.243 0.020 & 9.971 0.008 & 9.575 0.020 & \nodata \nodata & 9.324 0.009 & 9.328 0.009 & 9.322 0.008 \\
BH Peg       & \nodata \nodata & 10.899 0.005 & 10.426 0.005 & \nodata \nodata & 9.813 0.005 & \nodata \nodata & \nodata \nodata & \nodata \nodata & 9.000 0.008 & 8.979 0.008 \\
DH Peg       & 9.993 0.020 & 9.800 0.010 & 9.520 0.010 & \nodata \nodata & 9.113 0.010 & 8.806 0.020 & \nodata \nodata & 8.625 0.009 & 8.594 0.009 & 8.600 0.009 \\
RU Psc       & 10.534 0.020 & 10.458 0.004 & 10.162 0.004 & \nodata \nodata & 9.729 0.004 & \nodata \nodata & \nodata \nodata & \nodata \nodata & 9.088 0.009 & 9.075 0.008 \\
BB Pup       & 12.764 0.020 & 12.586 0.007 & 12.159 0.007 & 11.887 0.020 & 11.602 0.007 & 11.201 0.020 & 10.993 0.013 & 10.883 0.009 & 10.875 0.011 & 10.873 0.011 \\
HK Pup       & \nodata \nodata & 11.761 0.005 & 11.312 0.005 & \nodata \nodata & 10.707 0.005 & \nodata \nodata & \nodata \nodata & \nodata \nodata & 9.880 0.008 & 9.851 0.008 \\
RU Scl       & \nodata \nodata & 10.555 0.004 & 10.238 0.004 & \nodata \nodata & 9.805 0.004 & \nodata \nodata & \nodata \nodata & \nodata \nodata & 9.172 0.009 & 9.148 0.009 \\
SV Scl       & \nodata \nodata & 11.579 0.004 & 11.368 0.004 & \nodata \nodata & 11.007 0.004 & \nodata \nodata & \nodata \nodata & \nodata \nodata & 10.506 0.009 & 10.502 0.009 \\
AN Ser       & \nodata \nodata & 11.321 0.008 & 10.935 0.008 & \nodata \nodata & 10.446 0.008 & \nodata \nodata & \nodata \nodata & \nodata \nodata & 9.799 0.008 & 9.790 0.009 \\
AP Ser       & \nodata \nodata & 11.368 0.008 & 11.129 0.008 & \nodata \nodata & 10.765 0.008 & \nodata \nodata & \nodata \nodata & \nodata \nodata & 10.212 0.008 & 10.202 0.009 \\
T Sex       & 10.440 0.020 & 10.294 0.008 & 10.032 0.008 & 9.869 0.020 & 9.673 0.008 & 9.325 0.020 & \nodata \nodata & 9.157 0.009 & 9.141 0.009 & 9.119 0.008 \\
V0440 Sgr       & 10.847 0.020 & 10.703 0.008 & 10.312 0.008 & 10.103 0.020 & 9.805 0.008 & \nodata \nodata & \nodata \nodata & \nodata \nodata & 9.038 0.009 & 9.022 0.008 \\
V0675 Sgr       & \nodata \nodata & 10.706 0.007 & 10.298 0.007 & \nodata \nodata & 9.720 0.007 & \nodata \nodata & \nodata \nodata & \nodata \nodata & 8.954 0.009 & 8.929 0.009 \\
MT Tel       & \nodata \nodata & 9.188 0.007 & 8.966 0.007 & \nodata \nodata & 8.608 0.007 & \nodata \nodata & \nodata \nodata & \nodata \nodata & 8.087 0.008 & 8.073 0.009 \\
AM Tuc       & \nodata \nodata & 11.918 0.006 & 11.626 0.006 & \nodata \nodata & 11.188 0.006 & \nodata \nodata & \nodata \nodata & \nodata \nodata & 10.600 0.009 & 10.564 0.009 \\
AB UMa       & \nodata \nodata & 11.359 0.009 & 10.912 0.009 & \nodata \nodata & 10.342 0.009 & \nodata \nodata & \nodata \nodata & \nodata \nodata & 9.596 0.009 & 9.586 0.008 \\
RV UMa       & \nodata \nodata & 10.979 0.021 & 10.711 0.021 & \nodata \nodata & 10.336 0.021 & \nodata \nodata & \nodata \nodata & \nodata \nodata & 9.755 0.009 & 9.740 0.009 \\
SX UMa       & \nodata \nodata & 11.040 0.010 & 10.848 0.010 & \nodata \nodata & 10.532 0.010 & \nodata \nodata & \nodata \nodata & \nodata \nodata & 10.076 0.009 & 10.065 0.009 \\
TU UMa       & 10.246 0.020 & 10.156 0.007 & 9.816 0.007 & 9.582 0.020 & 9.319 0.007 & 8.899 0.020 & 8.717 0.013 & 8.642 0.009 & 8.619 0.009 & 8.605 0.009 \\
UU Vir       & 10.986 0.020 & 10.868 0.007 & 10.561 0.007 & 10.348 0.020 & 10.118 0.007 & 9.723 0.020 & \nodata \nodata & 9.496 0.009 & 9.486 0.008 & 9.480 0.008 \\

\enddata
\end{deluxetable*}

\begin{longrotatetable}
\begin{deluxetable*}{l r r r r r r r r r r }
\tabletypesize{\tiny}
\tablecaption{Amplitudes, minimum magnitude and rise times from \gloess{} light curves \label{tab:lcpars}}
\tablewidth{0pt}
\tablehead{
Name & \multicolumn{1}{c}{$U$} &  \multicolumn{1}{c}{$B$} & \multicolumn{1}{c}{$V$} & \multicolumn{1}{c}{$R_C$} & \multicolumn{1}{c}{$I_C$} & \multicolumn{1}{c}{$J$} & \multicolumn{1}{c}{$H$} & \multicolumn{1}{c}{$K_s$} & \multicolumn{1}{c}{[3.6]} & \multicolumn{1}{c}{[4.5]}
}
\startdata
SW And &   1.265  10.823   0.191 &   1.264  10.630   0.176 &   0.930  10.096   0.164 &   0.740   9.757   0.191 &   0.578   9.457   0.156 &   0.406   8.998   0.168 &   0.312   8.802   0.332 &   0.304   8.725   0.344 &   0.280   8.680   0.387 &   0.279   8.665   0.406 \\ 
XX And & \nodata \nodata \nodata &   1.286  11.584   0.188 &   0.990  11.117   0.184 & \nodata \nodata \nodata &   0.651  10.468   0.172 & \nodata \nodata \nodata & \nodata \nodata \nodata & \nodata \nodata \nodata &   0.296   9.616   0.441 &   0.278   9.575   0.391 \\ 
WY Ant &   1.077  11.685   0.164 &   1.172  11.663   0.152 &   0.910  11.209   0.152 &   0.751  10.918   0.148 &   0.595  10.596   0.148 &   0.414  10.155   0.137 &   0.296   9.902   0.375 &   0.311   9.807   0.359 &   0.269   9.753   0.453 &   0.273   9.740   0.410 \\ 
X Ari &   1.157  10.724   0.297 &   1.209  10.526   0.141 &   0.955   9.948   0.145 &   0.807   9.576   0.137 &   0.632   9.162   0.133 &   0.435   8.544   0.148 &   0.294   8.234   0.375 &   0.297   8.127   0.418 &   0.310   8.112   0.441 &   0.323   8.099   0.395 \\ 
AE Boo & \nodata \nodata \nodata &   0.496  11.141   0.453 &   0.391  10.844   0.449 & \nodata \nodata \nodata &   0.231  10.373   0.438 & \nodata \nodata \nodata & \nodata \nodata \nodata & \nodata \nodata \nodata &   0.107   9.816   0.504 &   0.106   9.801   0.520 \\ 
ST Boo & \nodata \nodata \nodata &   1.346  11.852   0.293 &   1.039  11.394   0.180 & \nodata \nodata \nodata &   0.684  10.811   0.176 & \nodata \nodata \nodata & \nodata \nodata \nodata & \nodata \nodata \nodata &   0.346  10.080   0.207 &   0.335  10.058   0.250 \\ 
TV Boo &   0.584  11.496   0.313 &   0.716  11.519   0.309 &   0.587  11.271   0.305 &   0.432  11.037   0.230 &   0.400  10.835   0.309 &   0.217  10.410   0.414 & \nodata \nodata \nodata &   0.125  10.258   0.391 &   0.134  10.266   0.348 &   0.137  10.253   0.355 \\ 
UY Boo & \nodata \nodata \nodata &   1.276  11.757   0.152 &   1.013  11.318   0.148 & \nodata \nodata \nodata &   0.673  10.713   0.148 & \nodata \nodata \nodata & \nodata \nodata \nodata & \nodata \nodata \nodata &   0.242   9.886   0.461 &   0.262   9.879   0.465 \\ 
ST CVn & \nodata \nodata \nodata &   0.549  11.871   0.422 &   0.418  11.552   0.410 & \nodata \nodata \nodata &   0.260  11.084   0.473 & \nodata \nodata \nodata & \nodata \nodata \nodata & \nodata \nodata \nodata &   0.120  10.504   0.578 &   0.124  10.479   0.504 \\ 
UY Cam & \nodata \nodata \nodata &   0.461  11.931   0.422 &   0.356  11.692   0.465 & \nodata \nodata \nodata &   0.195  11.308   0.414 & \nodata \nodata \nodata & \nodata \nodata \nodata & \nodata \nodata \nodata &   0.054  10.806   0.316 &   0.074  10.799   0.926 \\ 
YZ Cap &   0.598  12.068   0.402 &   0.640  11.885   0.484 &   0.489  11.547   0.402 &   0.407  11.333   0.391 &   0.314  11.055   0.387 & \nodata \nodata \nodata & \nodata \nodata \nodata & \nodata \nodata \nodata &   0.094  10.391   0.457 &   0.098  10.379   0.559 \\ 
RZ Cep &   0.619  10.460   0.406 &   0.693  10.277   0.445 &   0.540   9.673   0.406 & \nodata \nodata \nodata &   0.346   8.920   0.391 & \nodata \nodata \nodata & \nodata \nodata \nodata & \nodata \nodata \nodata &   0.115   7.935   0.379 &   0.105   7.920   0.348 \\ 
RR Cet &   1.112  10.606   0.148 &   1.196  10.552   0.145 &   0.919  10.102   0.145 &   0.755   9.813   0.152 &   0.598   9.511   0.145 &   0.453   9.069   0.168 & \nodata \nodata \nodata &   0.280   8.711   0.477 &   0.276   8.690   0.383 &   0.278   8.682   0.441 \\ 
CU Com & \nodata \nodata \nodata &   0.652  14.008   0.395 &   0.447  13.595   0.398 & \nodata \nodata \nodata &   0.298  13.027   0.324 & \nodata \nodata \nodata & \nodata \nodata \nodata & \nodata \nodata \nodata &   0.128  12.332   0.441 &   0.128  12.317   0.430 \\ 
RV CrB & \nodata \nodata \nodata &   0.687  11.982   0.320 &   0.526  11.658   0.320 & \nodata \nodata \nodata &   0.326  11.182   0.305 & \nodata \nodata \nodata & \nodata \nodata \nodata & \nodata \nodata \nodata &   0.105  10.556   0.301 &   0.129  10.538   0.297 \\ 
W Crt &   1.731  12.726   0.102 &   1.667  12.541   0.141 &   1.281  12.058   0.141 &   1.025  11.765   0.164 &   0.818  11.456   0.129 &   0.549  11.066   0.148 &   0.375  10.871   0.195 &   0.339  10.741   0.195 &   0.226  10.641   0.480 &   0.255  10.655   0.277 \\ 
UY Cyg & \nodata \nodata \nodata &   1.155  11.991   0.160 &   0.851  11.464   0.156 & \nodata \nodata \nodata &   0.538  10.766   0.164 & \nodata \nodata \nodata & \nodata \nodata \nodata & \nodata \nodata \nodata &   0.283   9.905   0.379 &   0.276   9.879   0.391 \\ 
XZ Cyg & \nodata \nodata \nodata &   1.599  10.484   0.297 &   1.242  10.074   0.262 & \nodata \nodata \nodata &   0.802   9.517   0.203 & \nodata \nodata \nodata & \nodata \nodata \nodata & \nodata \nodata \nodata &   0.224   8.799   0.293 &   0.237   8.787   0.324 \\ 
DX Del & \nodata \nodata \nodata &   0.948  10.769   0.219 &   0.706  10.239   0.223 & \nodata \nodata \nodata &   0.440   9.603   0.215 &   0.228   9.130   0.227 &   0.259   8.984   0.367 &   0.261   8.885   0.383 &   0.248   8.816   0.445 &   0.257   8.806   0.441 \\ 
SU Dra &   1.138  10.632   0.184 &   1.282  10.669   0.172 &   0.969  10.209   0.164 &   0.784   9.899   0.176 &   0.645   9.605   0.152 &   0.460   9.165   0.148 & \nodata \nodata \nodata &   0.300   8.837   0.348 &   0.287   8.800   0.395 &   0.290   8.782   0.391 \\ 
SW Dra &   1.261  11.373   0.172 &   1.210  11.332   0.164 &   0.919  10.874   0.152 &   0.733  10.570   0.160 &   0.592  10.268   0.145 & \nodata \nodata \nodata & \nodata \nodata \nodata & \nodata \nodata \nodata &   0.276   9.494   0.426 &   0.280   9.480   0.430 \\ 
CS Eri & \nodata \nodata \nodata &   0.649   9.567   0.441 &   0.488   9.247   0.422 & \nodata \nodata \nodata &   0.291   8.800   0.391 & \nodata \nodata \nodata & \nodata \nodata \nodata & \nodata \nodata \nodata &   0.115   8.185   0.473 &   0.107   8.170   0.461 \\ 
RX Eri &   1.057  10.643   0.176 &   1.150  10.575   0.176 &   0.886  10.062   0.176 &   0.704   9.736   0.164 &   0.569   9.406   0.164 & \nodata \nodata \nodata & \nodata \nodata \nodata &   0.303   8.565   0.422 &   0.279   8.530   0.402 &   0.283   8.507   0.430 \\ 
SV Eri & \nodata \nodata \nodata &   0.794  10.696   0.305 &   0.610  10.215   0.316 & \nodata \nodata \nodata &   0.401   9.562   0.289 & \nodata \nodata \nodata & \nodata \nodata \nodata & \nodata \nodata \nodata &   0.209   8.687   0.359 &   0.196   8.664   0.324 \\ 
RR Gem &   1.702  12.614   0.172 &   1.585  12.374   0.180 &   1.183  11.854   0.148 &   0.968  11.529   0.152 &   0.742  11.217   0.148 &   0.476  10.783   0.121 & \nodata \nodata \nodata &   0.300  10.415   0.320 &   0.316  10.457   0.180 &   0.308  10.433   0.297 \\ 
TW Her & \nodata \nodata \nodata &   1.688  12.278   0.137 &   1.302  11.792   0.133 & \nodata \nodata \nodata &   0.829  11.185   0.125 & \nodata \nodata \nodata & \nodata \nodata \nodata & \nodata \nodata \nodata &   0.310  10.449   0.371 &   0.351  10.445   0.176 \\ 
VX Her & \nodata \nodata \nodata &   1.618  11.599   0.109 &   1.250  11.152   0.113 & \nodata \nodata \nodata &   0.847  10.589   0.109 & \nodata \nodata \nodata & \nodata \nodata \nodata & \nodata \nodata \nodata &   0.337   9.803   0.180 &   0.323   9.788   0.172 \\ 
SV Hya &   1.366  11.447   0.145 &   1.571  11.433   0.141 &   1.241  10.996   0.141 & \nodata \nodata \nodata &   0.837  10.399   0.137 & \nodata \nodata \nodata & \nodata \nodata \nodata & \nodata \nodata \nodata &   0.291   9.560   0.207 &   0.272   9.531   0.211 \\ 
V Ind & \nodata \nodata \nodata &   1.358  10.781   0.160 &   1.072  10.370   0.164 &   0.867  10.098   0.141 &   0.694   9.794   0.133 & \nodata \nodata \nodata & \nodata \nodata \nodata & \nodata \nodata \nodata &   0.283   9.043   0.375 &   0.285   9.027   0.363 \\ 
BX Leo & \nodata \nodata \nodata &   0.594  12.158   0.508 &   0.467  11.838   0.426 & \nodata \nodata \nodata &   0.302  11.376   0.469 & \nodata \nodata \nodata & \nodata \nodata \nodata & \nodata \nodata \nodata &   0.107  10.733   0.438 &   0.123  10.729   0.512 \\ 
RR Leo &   1.608  11.719   0.145 &   1.644  11.647   0.148 &   1.293  11.212   0.148 &   1.053  10.929   0.137 &   0.859  10.629   0.141 &   0.577  10.217   0.168 & \nodata \nodata \nodata &   0.322   9.878   0.285 &   0.323   9.868   0.172 &   0.332   9.845   0.164 \\ 
TT Lyn &   0.829  10.640   0.191 &   0.905  10.596   0.184 &   0.688  10.154   0.176 &   0.565   9.870   0.180 &   0.450   9.549   0.180 &   0.334   9.094   0.188 & \nodata \nodata \nodata &   0.223   8.753   0.457 &   0.237   8.755   0.422 &   0.253   8.732   0.422 \\ 
RR Lyr &   1.044   8.579   0.172 &   1.111   8.546   0.168 &   0.860   8.072   0.160 & \nodata \nodata \nodata &   0.584   7.502   0.172 &   0.294   6.933   0.188 &   0.238   6.740   0.520 &   0.230   6.637   0.488 &   0.219   6.614   0.434 &   0.216   6.605   0.418 \\ 
RV Oct & \nodata \nodata \nodata &   1.459  12.013   0.148 &   1.131  11.442   0.148 &   0.912  11.085   0.152 &   0.727  10.681   0.148 &   0.493  10.146   0.133 &   0.315   9.830   0.188 &   0.304   9.701   0.402 &   0.283   9.683   0.355 &   0.349   9.729   0.398 \\ 
UV Oct & \nodata \nodata \nodata &   1.289  10.336   0.160 &   0.994   9.840   0.152 & \nodata \nodata \nodata &   0.642   9.203   0.152 & \nodata \nodata \nodata & \nodata \nodata \nodata & \nodata \nodata \nodata &   0.249   8.349   0.426 &   0.248   8.341   0.422 \\ 
AV Peg &   1.409  11.686   0.195 &   1.360  11.454   0.184 &   0.991  10.899   0.172 &   0.808  10.596   0.195 &   0.624  10.269   0.160 &   0.373   9.789   0.219 & \nodata \nodata \nodata &   0.275   9.504   0.422 &   0.283   9.526   0.375 &   0.280   9.513   0.332 \\ 
BH Peg & \nodata \nodata \nodata &   0.845  11.267   0.195 &   0.636  10.729   0.199 & \nodata \nodata \nodata &   0.416  10.038   0.203 & \nodata \nodata \nodata & \nodata \nodata \nodata & \nodata \nodata \nodata &   0.238   9.157   0.453 &   0.238   9.133   0.469 \\ 
DH Peg &   0.565  10.283   0.402 &   0.629  10.137   0.434 &   0.503   9.783   0.453 & \nodata \nodata \nodata &   0.305   9.268   0.391 &   0.222   8.928   0.465 & \nodata \nodata \nodata &   0.113   8.688   0.551 &   0.088   8.646   0.586 &   0.139   8.671   0.574 \\ 
RU Psc &   0.542  10.830   0.430 &   0.582  10.756   0.453 &   0.465  10.399   0.477 & \nodata \nodata \nodata &   0.282   9.862   0.441 & \nodata \nodata \nodata & \nodata \nodata \nodata & \nodata \nodata \nodata &   0.065   9.115   0.352 &   0.061   9.105   0.504 \\ 
BB Pup &   1.311  13.318   0.172 &   1.300  13.141   0.164 &   0.981  12.586   0.148 &   0.782  12.251   0.148 &   0.623  11.912   0.145 &   0.406  11.456   0.191 &   0.350  11.240   0.508 &   0.333  11.114   0.430 &   0.278  11.066   0.332 &   0.280  11.059   0.461 \\ 
HK Pup & \nodata \nodata \nodata &   0.885  12.168   0.219 &   0.670  11.632   0.207 & \nodata \nodata \nodata &   0.430  10.938   0.227 & \nodata \nodata \nodata & \nodata \nodata \nodata & \nodata \nodata \nodata &   0.252  10.044   0.438 &   0.258  10.015   0.434 \\ 
RU Scl & \nodata \nodata \nodata &   1.567  11.208   0.113 &   1.160  10.685   0.125 & \nodata \nodata \nodata &   0.738  10.116   0.105 & \nodata \nodata \nodata & \nodata \nodata \nodata & \nodata \nodata \nodata &   0.283   9.368   0.262 &   0.267   9.333   0.313 \\ 
SV Scl & \nodata \nodata \nodata &   0.642  11.931   0.445 &   0.514  11.646   0.449 & \nodata \nodata \nodata &   0.292  11.155   0.336 & \nodata \nodata \nodata & \nodata \nodata \nodata & \nodata \nodata \nodata &   0.100  10.556   0.434 &   0.102  10.545   0.445 \\ 
AN Ser & \nodata \nodata \nodata &   1.373  11.950   0.195 &   1.000  11.401   0.180 & \nodata \nodata \nodata &   0.625  10.768   0.168 & \nodata \nodata \nodata & \nodata \nodata \nodata & \nodata \nodata \nodata &   0.293   9.997   0.410 &   0.299   9.991   0.391 \\ 
AP Ser & \nodata \nodata \nodata &   0.646  11.726   0.453 &   0.478  11.386   0.469 & \nodata \nodata \nodata &   0.316  10.934   0.441 & \nodata \nodata \nodata & \nodata \nodata \nodata & \nodata \nodata \nodata &   0.108  10.270   0.449 &   0.123  10.272   0.500 \\ 
T Sex &   0.469  10.693   0.438 &   0.534  10.596   0.465 &   0.411  10.262   0.465 &   0.330  10.051   0.441 &   0.257   9.815   0.465 &   0.201   9.427   0.500 & \nodata \nodata \nodata &   0.090   9.208   0.500 &   0.082   9.183   0.523 &   0.087   9.165   0.492 \\ 
V0440 Sgr &   1.442  11.376   0.125 &   1.551  11.300   0.133 &   1.205  10.773   0.137 &   0.985  10.491   0.133 &   0.790  10.136   0.129 & \nodata \nodata \nodata & \nodata \nodata \nodata & \nodata \nodata \nodata &   0.322   9.262   0.383 &   0.304   9.231   0.176 \\ 
V0675 Sgr & \nodata \nodata \nodata &   1.224  11.192   0.148 &   0.942  10.681   0.152 & \nodata \nodata \nodata &   0.621  10.008   0.152 & \nodata \nodata \nodata & \nodata \nodata \nodata & \nodata \nodata \nodata &   0.286   9.157   0.391 &   0.286   9.131   0.398 \\ 
MT Tel & \nodata \nodata \nodata &   0.668   9.529   0.375 &   0.531   9.224   0.352 & \nodata \nodata \nodata &   0.334   8.775   0.367 & \nodata \nodata \nodata & \nodata \nodata \nodata & \nodata \nodata \nodata &   0.081   8.124   0.266 &   0.087   8.116   0.242 \\ 
AM Tuc & \nodata \nodata \nodata &   0.577  12.231   0.434 &   0.441  11.860   0.441 & \nodata \nodata \nodata &   0.274  11.334   0.457 & \nodata \nodata \nodata & \nodata \nodata \nodata & \nodata \nodata \nodata &   0.124  10.653   0.543 &   0.138  10.633   0.461 \\ 
AB UMa & \nodata \nodata \nodata &   0.550  11.615   0.281 &   0.395  11.106   0.281 & \nodata \nodata \nodata &   0.266  10.493   0.301 & \nodata \nodata \nodata & \nodata \nodata \nodata & \nodata \nodata \nodata &   0.173   9.705   0.523 &   0.178   9.694   0.512 \\ 
RV UMa & \nodata \nodata \nodata &   1.484  11.614   0.254 &   1.124  11.171   0.246 & \nodata \nodata \nodata &   0.691  10.639   0.230 & \nodata \nodata \nodata & \nodata \nodata \nodata & \nodata \nodata \nodata &   0.331   9.955   0.188 &   0.338   9.949   0.199 \\ 
SX UMa & \nodata \nodata \nodata &   0.655  11.383   0.402 &   0.500  11.108   0.398 & \nodata \nodata \nodata &   0.299  10.687   0.340 & \nodata \nodata \nodata & \nodata \nodata \nodata & \nodata \nodata \nodata &   0.107  10.142   0.488 &   0.109  10.125   0.512 \\ 
TU UMa &   1.113  10.690   0.152 &   1.237  10.666   0.152 &   0.949  10.213   0.145 &   0.762   9.914   0.160 &   0.610   9.611   0.145 &   0.425   9.146   0.129 &   0.317   8.929   0.395 &   0.298   8.852   0.457 &   0.297   8.832   0.383 &   0.286   8.812   0.418 \\ 
UU Vir &   1.502  11.638   0.148 &   1.487  11.489   0.145 &   1.133  11.030   0.145 &   0.933  10.770   0.137 &   0.734  10.460   0.125 &   0.391   9.980   0.293 & \nodata \nodata \nodata &   0.326   9.722   0.449 &   0.313   9.710   0.332 &   0.302   9.697   0.195 \\ 

\enddata
\end{deluxetable*}
\end{longrotatetable}
\clearpage
\bibliographystyle{aasjournal}
\bibliography{ms} 

\appendix
\section{Image Processing} \label{app:img_prox}
The following section discusses the data reduction procedure under the following prescription:\begin{equation}\scriptsize{
 I^*(x,y)_{i} = B(x,y) + D(x,y)(E_i) + (G)F(x,y)_i(S(x,y)_i(E(x,y) + E_i))}
\end{equation} where  $I^*(x,y)_{i}$ is the ideal (linear response) raw image and $G$ is the average gain (ADU/photon) of the detector.

In the case of a nonlinear detector the actual raw image, $I(x,y)_{i}$,  approximately becomes
 \begin{equation}
 I(x,y)_{i} =  I^*(x,y)_{i}(1 - L(x,y)(I^*(x,y)_{i}) )
 \end{equation}
 where $L(x,y)_{i}$ is the nonlinearity of the pixel.  
Typically this is a small number, resulting in a reduction in counts by only a few per cent.  
The ultimate goal is to produce an image that contains the source count rate $S(x,y)_{i}$  of the target being imaged:
\begin{equation} \label{eq:finalredux}
\scriptsize{S(x,y)_i = \frac{I(x,y)_{i}(1+L(x,y)I(x,y)_{i} )-B(x,y)-D(x,y)(E_i) }{ (G)F(x,y)_i((E(x,y) + E_i)) }  }
\end{equation}
where each component of Equation \ref{eq:finalredux} is described in one of the following subsections.

\subsection{Bias and Dark Frames: $B(x,y)$ and $D(x,y)$}

A bias frame is the intrinsic count level on the detector and it can be single-valued or have a spatial structure. In our system the bias pattern has noticeable column structure, which actually changes when the  shutter is actively held open versus closed.  The origin of the effect is not understood\footnote{The camera has since been repaired by E2V for a shutter power problem, which also fixed the bias problem.}; however, calibrations can be taken with the shutter open by placing a ``dark" blocking filter in the light path.  We typically obtain thirty 5 s ``light" frames with the dark filter in place. To construct the 2D bias these frames were averaged together (with sigma-clipping to reject cosmic rays or other outliers).

A dark frame is a map of how many counts each pixel produces due to its own heat or electrical characteristics.  Our dark frame is produced in much the same way as our bias. 
We typically take five 605 s ``light" frames with the dark filter in place.  
To construct the 2D dark these frames are averaged together (with sigma-clipping to reject cosmic rays or other outliers). 
The reason for taking long dark frames is because dark counts are typically low and short exposures are dominated by read noise, so getting enough dark signal requires longer exposures. 
One could average a larger number of shorter dark frames, but that requires more overhead in the form of read-out time and data transfer.  


The average bias is subtracted from the average dark, which removes the bias component and leaves only the dark rate for a 600 s exposure (605 s - 5 s) since the average bias contains 5 s of dark counts.  The master dark ($D(x,y)$) is then normalized to counts per second.  The scaled master dark frame is then subtracted from the original bias to create the dark-subtracted master bias frame, $B(x,y)$.   

\subsection{Exposure correction image: $E(x,y)$}

\begin{figure}
\begin{centering}
\includegraphics[width=\columnwidth]{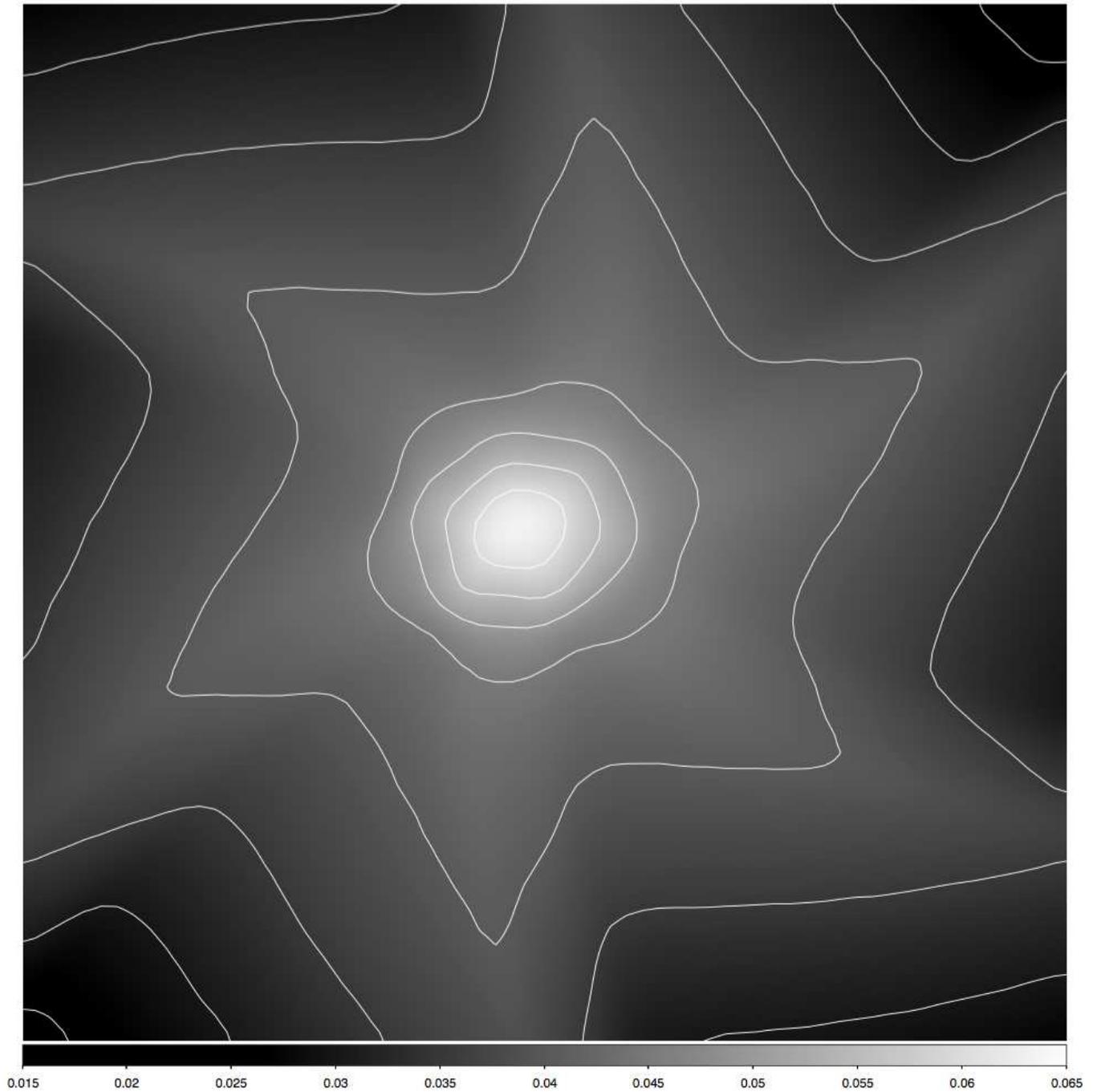}
\caption{The TMMT shutter correction image.  The six blades of the iris shutter open outwards and close in a finite amount of time, resulting in an uneven illumination pattern.  The resulting map of exposure time varies radially from 0.07 to 0.02 s. Starting from the center, the contours represent: 60, 55, 50, 45, 40, 35, 30, and 25 additional milliseconds to the original exposure time.}
\label{shuttertcor}
\end{centering}
\end{figure}

The camera uses a 63 mm Melles-Griot iris shutter, which has a finite opening and closing time, 
 meaning that the center of the CCD is exposed slightly longer than the edges.
As a result, the true exposure time is the sum of the set exposure time $E_i$ and the inherent shutter correction $E(x,y)$. 
To find the shutter correction, pairs of exposures (short and long) at various exposure times were taken while viewing a flat-field screen.  
Each image pair was processed up to the current point, 
\begin{equation} 
\scriptsize{A(x,y)_i = \frac{I(x,y)_{i}(1+L(x,y)I(x,y)_{i} )-B(x,y)-D(x,y)(E_i) }{ (G)F(x,y)_i }}
\end{equation}
and the shutter correction constructed from the following relation: 
\begin{equation}
E(x,y)  = \frac{A_1E_2 - A_2E_1}{A_2 - A_1}.
\end{equation} 
Multiple pairs of exposure times and flat-field intensities were used to construct an average shutter correction. The best result is obtained by using a uniform screen bright enough to get a good signal-to-noise ratio, with a short shutter time ($E_1$) just longer than the open/close time of the shutter; in this case $E_1=0.1 s$.  A corresponding long shutter time ($E_2$) is enough to avoid saturation at the same light level as ($E_1$).  Since the shutter correction is a relatively low-order smooth feature the image was smoothed using a wavelet transform to remove the pixel noise; see Figure \ref{shuttertcor}.

\subsection{Flats Frames and Linearity: $F(x,y)$ and $L(x,y)$}
Flat-field frames were taken by using an 18" Alnitak\footnote{Optec Inc. \url{http://www.optecinc.com}} 
 electroluminescent flat-field panel.  
The illumination was adjusted for each filter to reach 30,000 ADU in 30 s.  
The relatively long exposure time was designed to minimize the impact of the shutter correction,
 which at 30 s was already less than a 0.2\% correction.  
To reduce the effect of stray light, exposures were taken with the panel on and off; 
 the images were processed and then differenced to produce a flat field.   
To determine the linearity a series of exposures of increasing time were taken of the flat-field screen.  
The image counts versus exposure time for each pixel were fit with a second-order polynomial.  
The first term in the polynomial is the linear response component and the linearity term 
 is the ratio of the second coefficient to the first.  
For this camera there is no significant pixel-to-pixel variation, and a constant value of 
{$10^{-6}$}
was adopted for the entire array corresponding to a 5\% correction at 50,000 ADU.   
To place the final image in more meaningful (Poisson) units the image was divided by the average gain; 
 $G$=1.4 in this case as adopted from the manufacturer and confirmed from measurements of variance versus flux.  
%
\section{Notes on Individual Stars} \label{app:indstars}
In this appendix, we organize discussions relating to specific stars in our sample,
 incorporating our literature search and analysis.
More specifically, we provide rationales regarding the adopted parameters for the stars when there are disagreements in the literature (i.e., the values given in Table \ref{tab:sample}), 
 any notes regarding the behavior of the star, 
 special techniques required to phase data sets widely separated in time,
 and any other individual star concerns. 
By separating these discussions from the main body of the paper we hope to 
 be as streamlined as possible in the example schematic presented in the main text, 
 while also providing as complete documentation of our efforts as possible. 

Unless otherwise noted, all 55 stars have data from the following sources:
 (i) \BVI{} multiphase optical observations from the TMMT 
 (Section \ref{sec:tmmt} in the main text), 
 (ii) one \JHK{} phase point from 2MASS (Section \ref{sec:2massdata}), and
 (iii) \LM{} multiphase data from \spitzer{} (Section \ref{sec:spitzerdata}) and \wise{} (Section \ref{sec:wisedata}). 
Filter transformations or other modifications required for the entire body of a study 
 were given in the appropriate subsections of Section \ref{sec:data},
 with star-by-star specifics reserved for the following subsections.

\subsection{SW And} \label{sw_and}
SW And shows the Blazhko effect with $P_{Bl} = 36.8$ days \citep{smith_1995}. 
Supplementary data came from the following sources:
$B$,$V$,$I$ from \citet{Barnes_1992},
$B$,$V$,$I$ from \citet{Barcza_2014},
$B$,$V$,$K$ from \citet{jones_1992},
$B$,$I$,$R$,$U$,$V$,$J$,$K$ from \citet{liu_1989}, and 
$J$,$H$,$K_s$ from \citet{Barnes_1992}. 

\subsection{XX And} \label{xx_and}
XX And is a non-Blazhko RRab.
Supplementary data in \JHK{} from \citet{fernley_1993}. 

\subsection{WY Ant} \label{xy_ant}
 WY Ant is a non-Blazhko RRab.
 Supplementary data come from the following sources:
  $U$, $B$, $V$, $R$, $I$ from \citet{skillen_1993},
  $V$ from ASAS, and
  \JHK{} from \citet{skillen_1993jhk}.
\subsection{X Ari} \label{x_ari}
  X Ari is a non-Blazhko RRab.
  Supplementary data come from the following sources:
   $V$ from ASAS and $B$, $V$, $R$, $I$, $J$, $H$, and $K$ from \citet{fernley_1989}. 

\subsection{AE Boo} \label{ae_boo}
  AE Boo is a non-Blazhko RRc.
  There are no supplementary data for this star.

\subsection{ST Boo} \label{st_boo}
  ST Boo is a RRab showing the Blazhko effect $P_{Bl} = 284.09$ days \citep{smith_1995}.
  Owing to this long period, we only see hints { of two distinct amplitudes in our light curve.}
  Supplementary \JHK{} data come from \citet{fernley_1993}.

\subsection{TV Boo} \label{tv_boo}
  TV Boo is a RRc showing the Blazhko effect \citep{smith_1995}.
  \citet{skarka_2013b} showed the star to have { two Blazhko periods of }
    $P_{1} = 9.737$ days and $P_{2} = 21.5$ days.
  In our light curve we see some slight indications of this behavior.
  Supplementary data come from the following sources:
    $U$, $B$, $V$, $R$, $I$, $J$, $K_s$ from \citet{liu_1989}, 
    \JHK{} from \citet{fernley_1993}, and 
    $U$, $B$, $V$ from \citet{Pac_1965_2}.

\subsection{UY Boo} \label{uy_boo}
  UY Boo is a RRab showing the Blazhko effect with $P_{Bl}$ = 171.8 days \citep{skarka_2014}.
 { Between $\phi$ = 0.1 and $\phi$ = 0.2 in our data, we see clearly see two distinct amplitudes for the light curve.} 
  Supplementary data in \JHK{} come from \citet{fernley_1993}.

\subsection{UY Cam} \label{uy_cam}
  UY Cam was only observed for one night with the TMMT. 
  The light curve is supplemented in $B$ and $V$ with data from \cite{Broglia1992}.  
  An offset of 0.03 mag and phase offset of $\phi$=0.02 were required to align the data to the modern epoch.  
  There appears to be some slight modulation of the amplitude that may account for the offset. 
  \cite{Zhou_2003} studied this star in much greater detail, and 
   as a result they determine this star to be an analog to a high-amplitude Delta Scuti (HADS), SX Phoenicis, and type RRc variable.  For the purposes of this paper we continue to classify it as type RRc and provide the photometry.  

\subsection{YZ Cap} \label{yz_cap}
  YZ Cap is a non-Blazhko RRc.
  Supplementary data in $U$, $B$, $V$, $R$, $I$ from \citet{cacciari_1987}.

\subsection{RZ Cep} \label{rz_cep}
  RZ Cep is a non-Blazhko RRc and the only RRc with an \hst~ parallax from \citet{benedict_2011}, in which two values are given for the parallax of RZ Cep: $\pi$ = 2.12 and $\pi$ = 2.54 mas, the former being the final and preferred adopted value (\cite{benedict_2011}, private communication).  New multi-wavelength PL relations suggest, however, that $\pi$ = 2.54 mas provides a better solution (Monson et al. in prep).  This is also consistent with results from \cite{Juna2013}, which find the absolute magnitude of RRc from statistical parallax to be $M_V = 0.590\pm0.103\pm0.014$ at [Fe/H]=-1.5 dex. For RZ Cep the {value of $M_V$ is 0.25 mag or 0.64 mag using 2.12 or 2.54 mas}, respectively (adopting $\langle V \rangle=9.398$ mag and $A_V$ of 0.78 mag).

Supplementary data for this star come from $U$, $B$, $V$ in \citet{Pac_1965_2} and \JHK{} from \citet{fernley_1993}.

The data presented for RZ Cep in \citet{feast_2008} appear to contain a discrepancy. 
While \citet{fernley_1989} have $E(B-V)$ = 0.25 mag, \citet{feast_2008} claim $E(B-V)$ = 0.078 mag.  
  \citet{benedict_2011} found $A_v$ $\sim$1.0 for stars along the line of sight to RZ Cep, suggesting the former is closer to correct. 
  \citet{feast_2008} describe using an average reddening between the Galactic extinction model of \cite{Drimmel_2003} and distance estimated using the PL(K) and M$_V$-\text{[Fe/H]} relations and iterating to find $E(B-V)$.  
  As pointed out in Sec \ref{sec:comp} their adopted mean $V$ magnitude for RZ Cep was transformed from the \hip{} H$_P$ magnitude using an adopted mean color correction for type RRc RRL. 
Since RZ Cep is highly reddened, this color correction would cause the distance modulus in the M$_V$-\text{[Fe/H]} relation to be overestimated and the resulting estimate of $E(B-V)$ to be underestimated.    

\subsection{RR Cet} \label{rr_cet}
  RR Cet is a non-Blazhko RRab. 
  Supplementary data come from \citet{liu_1989} in the 
   $U$, $B$, $V$, $R$, $I$, $J$, and $K_s$ bands.

\subsection{CU Com} \label{cu_com}
 CU Com is the only double-mode pulsator in our sample. 
  \citet{clementini_2000} { used \BVI{} photometry with an 11 year baseline, 
  finding that CU Com has to have periods $P_0$ = 0.5441641 days and $P_1$ = 0.4057605
  ($P_{1}/P_{0}$ = 0.7457).}
 The HJD data for the $B$ observations presented in \citet{clementini_2000} 
  and those available from the online catalog differ by 2231 days,
  which was rectified for the data in this study.   

\subsection{RV CrB} \label{rv_crb}
  RV CrB is a non-Blazhko RRc.
  Additional \JHK~data come from \citet{fernley_1993}.

\subsection{W Crt} \label{w_crt}
  W Crt is a non-Blazhko RRab.
  $U$,$B$,$V$,$R$,$I$ data were adopted from \cite{skillen_1993}, 
   but required a correction of $\delta M_{\lambda}$ = -0.04 mag.
  \JHK~data were adopted from \citet{skillen_1993jhk}.
 
\subsection{ST CVn} \label{ct_cvn}
  ST CVn is a non-Blazhko RRc.
  Additional \JHK{} data were adopted from \citet{fernley_1993}.

\subsection{UY Cyg} \label{xy_cyg}
  UY Cyg is a non-Blazhko RRab.
  Additional \JHK{} data were adopted from \citet{fernley_1993}.

\subsection{XZ Cyg} \label{xz_cyg}
  XZ Cyg is an RRab showing the Blazhko effect with $P = 57.3$ days with a time-varying period as identified in \citet{lacluyze_2004}.  
   Additional \JHK~data were adopted from \citet{fernley_1993}.
  Our TMMT light curve clearly shows the Blazhko effect from $\phi=0.8$ to $\phi=1.0$.
  To remedy problems in the smoothed light curve from the Blazhko effect, 
   the \gloess{} light curve sampled the data at 50 phase points and then the \gloess{} light curve was up-sampled to 200 phase points.
  We note that the mismatch in the MIR and NIR data points is likely due to being
   on different parts of the Blazhko phase. 
   
\subsection{DX Del} \label{dx_del}
  DX Del is a non-Blazhko RRab.
  Supplementary data are adopted from \citet{skillen_1989} in $V$,\JHK. 

\subsection{SU Dra} \label{su_dra}
  SU Dra is a non-Blazhko RRab. 
 Supplementary data are adopted from \citet{liu_1989} in 
  $U$, $B$, $V$, $R$, $I$, $J$ and $K_s$. 

\subsection{SW Dra} \label{sw_dra}
  SW Dra is a non-Blazhko RRab.
  Supplementary data in $U$, $B$, $V$, $R$, $I$ are from \citet{cacciari_1987}.

\subsection{CS Eri} \label{cs_eri}
  CS Eri is a non-Blazhko RRc.
  There are no supplementary data for this star.

\subsection{RX Eri} \label{rx_eri}
  RX Eri is a non-Blazhko RRab.
  Additional data are adopted from \citet{liu_1989} 
   in $U$, $B$, $V$, $R$, $I$, $J$, and $K_s$.

\subsection{SV Eri} \label{sv_eri}
  SV Eri is a non-Blazhko RRab
  Additional data in \JHK~come from \citet{fernley_1993}.

\subsection{RR Gem} \label{rr_gem}
  The light curve for RR Gem is classified as `stable or contradictory,'
   meaning that it does not always show a Blazhko modulation, but has at times
   \citep[][and references therein]{jurcsik_2009}.
  The most recent characterization is from \citet{jurcsik_2005},
    finding a short period, $P_{Bl} = 7.23$ day, 
    and a small 0.1 mag modulation.  
  Our data show a clear Blazhko effect over the full phase cycle. 
  Supplementary data are adopted from 
    \citet{liu_1989} in $U$, $B$, $V$, $R$, $I$, $J$, and $K_s$.
  
\subsection{TW Her} \label{tw_her}
  TW Her is a non-Blazhko RRab.
  No additional data were adopted.

\subsection{VX Her} \label{vx_her}
   VX Her is an RRab showing a very long-period Blazhko effect with
     $P_{Bl}=455.37$ days { (the Blazhko period was first published in \citet{wunder_1990}, but the associated data are much more easily accessed in the compilation of \citet{skarka_2013})}.
   Owing to this long period, we see no evidence of a Blazhko effect in our data.
   Supplementary data are adopted in \JHK{} from \citet{fernley_1993}.
   

\subsection{SV Hya} \label{sv_hya}
  SV Hya is an RRab showing a Blazhko effect with 
   $P_{Bl} = 63.29$ days \citep{smith_1995}.
  Visible discontinuities in our light curve can be attributed to this effect.   
  Supplementary data are adopted from \citet{fernley_1993} in \JHK{} and from \citet{Warren_1966} in $U$, $B$, $V$. 

To construct a light curve without a large discontinuity, 
   the data on MJD 57160 were shifted by -0.12 mag. 
  The data being presented are unaltered in the primary catalog, 
   but the modified data used to build the \gloess{} light curve 
   are available under the ID code 999 for $B$, $V$, $I$.  

\subsection{V Ind} \label{v_ind}
  V Ind is a non-Blazhko RRab. 
  We supplement the TMMT data with $B$, $V$, $R$, $I$ data from \citet{Clementini_1990} and $V$ data from ASAS.

\subsection{BX Leo} \label{bx_leo}
  BX Leo is a non-Blazhko RRc.
  No supplementary data are adopted for this star. 
  This star was unusually difficult to phase properly.
  Phasing of the \spitzer{} epochs required an offset of $\Delta \phi$=0.15.
  
\subsection{RR Leo} \label{rr_leo}
  RR Leo is a non-Blazhko RRab.
  Supplementary data are adopted from \citet{liu_1989} in 
  $U$, $B$, $V$, $R$, $I$, $J$ and $K$. 

\subsection{TT Lyn} \label{tt_lyn}
  TT Lyn is a non-Blazhko RRab.
  Supplementary data are adopted from \citet{liu_1989} in 
  $U$, $B$, $V$, $R$, $I$, $J$ and $K$ 
   and from \citet{Barnes_1992} in $B$, $V$, $I$.

\subsection{RR Lyr} \label{rr_lyr}
  RR Lyr is an RRab star that is well known to exhibit the
   Blazhko effect \citep{smith_1995}. 
  Moreover, RR Lyr has a time-variable Blazhko cycle 
    with a variation from $P_{Bl} = 38.8$ to $P_{Bl} = 40.8$ days
    \citep{kolenberg_2006}.
  RR Lyr is in the {\emph Kepler} field and its time variations
   are explored in detail within {\emph Kepler} {\bf by \citet{kolenberg_2011} and} 
   compared to an abundance of historical data by \citet{leborgne_2014}.
  Supplementary data in \JHK~are adopted 
   from \citet{sollima_2008} and \citet{fernley_1993}.

\subsection{RV Oct} \label{rv_oct}
  RV Oct is a non-Blazhko RRab.
  Supplementary data are adopted from \citet{skillen_1993} in $B$,$V$,$R$,$I$
   and \citet{skillen_1993jhk} in \JHK.
   
\subsection{UV Oct} \label{uv_oct}
  UV Oct is an RRab showing the Blazhko effect 
   with $P_{Bl} = 143.73$ days \citep{skarka_2013}.
  We see bifurcation in our light curve indicative
   of seeing the Blazhko effect.
  Supplementary data are adopted in $V$ from ASAS.

\subsection{AV Peg} \label{av_peg}
  AV Peg is a non-Blazhko RRab.
  Supplementary data are adopted from \citet{liu_1989} in 
   $U$,$B$,$V$,$R$,$I$,$J$, and $K$.

\subsection{BH Peg} \label{bh_peg}
  BH Peg is an RRab showing the Blazhko effect with 
   $P_{Bl} = 39.8$ days \citep{smith_1995}.
  We see no evidence for a Blazhko effect in our light curve.
  Supplementary data are adopted in \JHK{} from \citet{fernley_1993}.

\subsection{DH Peg} \label{dh_peg}
  DH Peg is a non-Blazhko RRc.
  Additional data are adopted from the following sources:
   \cite{fernley_1990} in $V$,$J$,$K_{UKIRT}$,
   \citet{Barcza_2014} in $B$,$V$,$I$, and 
   \citet{Pac_1965_2} in $U$,$V$,$B$.  
 There are no \emph{Spitzer} data for this star.   

\subsection{RU Psc} \label{ru_psc}
  RU Psc is an RRc showing the Blazhko effect 
   with $P_{Bl} = 28$ days \citep{smith_1995}. 
  We see evidence of amplitude modulation in our light curve.
  Supplementary data are obtained from \citet{Pac_1965_2} in $U$,$B$,$V$
   and \citet{fernley_1993} in \JHK. 
  Phasing of the \spitzer{} epochs required an offset of $\Delta \phi$=0.12.

\subsection{BB Pup} \label{bb_pup}
  BB Pup is a non-Blazhko RRab.
  Supplementary data from 
   \cite{skillen_1993} in $U$,$B$,$V$,$R$,$I$, 
   \cite{skillen_1993jhk} in \JHK, and 
   $V$ from ASAS.
  There are no \spitzer~ data for this star.

\subsection{HK Pup} \label{hk_pup}
  HK Pup is a non-Blazhko RRab.
  Supplementary data in $V$ from ASAS.

\subsection{RU Scl} \label{ru_scl}
  RU Scl is an RRab showing the Blazhko effect 
   with $P_{Bl}$ = 23.91 days \citep{skarka_2014}.
  We see some modulation in our light curve at maximum light.
  Supplementary data in \JHK{} are adopted from \citet{fernley_1993}.
  There are no \spitzer~ data for this field. 

\subsection{SV Scl} \label{sv_scl}
  SV Scl is a non-Blazhko RRc. 
  Supplementary data are adopted in \JHK{} from \citet{fernley_1993}
   and in $V$ from ASAS.

\subsection{AN Ser} \label{an_ser}
  AN Ser is a non-Blazhko RRb. 
  Supplementary data are adopted in \JHK{} from \citet{fernley_1993}
   and in $V$ from ASAS.

\subsection{AP Ser} \label{ap_ser}
  AP Ser is a non-Blazhko RRc. 
  There are no additional data for this star.

\subsection{T Sex} \label{t_sex}
  T Sex is a non-Blazhko RRc.
  Supplementary data are adopted from the following sources:
   \citet{Barnes_1992} in $B$,$V$,$I$,
   \citet{liu_1989} in $U$,$B$,$V$,$R$,$I$,$J$, and $K$, and 
   $V$ from ASAS.

\subsection{V0440 Sgr} \label{v0440_sgr}
  V0440 Sgr is a non-Blazhko RRab.
  Supplementary data are adopted from 
   \citet{cacciari_1987} in $U$,$B$,$V$,$R$,$I$ and 
   \citet{fernley_1993} in \JHK. 

\subsection{V0675 Sgr} \label{v0675_sgr}
  V0675 Sgr is a non-Blazhko RRab. 
  Supplementary data in \JHK{} are adopted from \citet{fernley_1993}.

\subsection{MT Tel} \label{mt_tel}
  MT Tel is a non-Blazhko RRc. 
  Additional $V$ data are adopted from ASAS.

\subsection{AM Tuc} \label{am_tuc}
  AM Tuc is a RRc that has a long-period Blazhko modulation 
   with $P_{Bl} = 1748.86$ days \citet{sf_2007}.
  This period is much longer than the span of our observations.
  No additional data are adopted for this star. 

\subsection{AB UMa} \label{ab_uma}
  AB UMa is a non-Blazhko RRab.
  No additional data are used for this star, although there is good agreement between the average found here and that of \cite{Piersimoni1993}.

\subsection{RV UMa} \label{rv_uma}
  RV UMa is an RRab showing the Blazhko effect \citep{smith_1995}.
  \citet{hurta_2008} note a time-variable Blazhko period 
    from $P_{Bl} = 89.9$ to $P_{Bl} = 90.63$ days.
  Supplementary data are adopted in \JHK{} from \citet{fernley_1993}.  
    
 We see evidence for modulation in our light curve.  
 In order to construct a \gloess{} light curve without a large discontinuity, 
  the amplitudes of the light curves from MJD 56786, 56793, and 56796  
  were scaled by a factor of 1.6 around the B,V,I means of 11.2, 10.9, 10.4, respectively.  
  The data are presented unaltered in the primary table, 
   but the scaled data are available under the ID code 999 for $B$, $V$, $I$.  
  
\subsection{SX UMa} \label{sx_uma}
  SX UMa is a non-Blazhko RRc.
  Supplementary data are adopted in \JHK{} from \citet{fernley_1993}.

\subsection{TU UMa} \label{tu_uma}
  TU UMa is an RRab. 
  As discussed in \citet[][and references therein]{Barnes_1992},
  \citet{fernley_1997},
   and \citet{Liska2016}, this might be in a binary system,
   which causes slight delays in the phasing of the { time of maximum light for the light curve.} 
  Supplementary data are adopted in
   $B$,$V$,$I$,\JHK{} from \citet{Barnes_1992} and 
   $U$,$B$,$V$,$I$,$R$,$J$ and $K$ from \citet{liu_1989}.

\subsection{UU Vir} \label{uu_vir}
  UU Vir is a non-Blazhko RRab.
  Supplementary data are adopted from the following sources:
   $U$,$B$,$V$,$I$,$R$,$J$ and $K$ from \citet{liu_1989},
   $V$ from ASAS, and
   \JHK{} from \citet{Barnes_1992}.

%
 
   
\end{document}